\UseRawInputEncoding

\documentclass[pra,twocolumn,showpacs,groupedaddress,superscriptaddress,nofootinbib,floatfix,preprintnumbers,longbibliography]{revtex4-1}

\usepackage{amssymb,amsmath,graphicx,color}
\usepackage{physics}
\usepackage{bm}
\usepackage[tight]{subfigure}
\usepackage[export]{adjustbox}
\usepackage{braket}
\usepackage[colorlinks=true,citecolor=blue,linkcolor=red]{hyperref}
\usepackage{multirow}
\usepackage{rotating,booktabs}
\usepackage[verbose]{placeins}
\usepackage{color}
\usepackage{ulem}

\begin{document}

\title{
Engineering an Effective Three-spin Hamiltonian in Trapped-ion Systems
\\
  for Applications in Quantum Simulation
}

\author{B\'{a}rbara Andrade}
\affiliation{
ICFO-Institut de Ciencies Fotoniques, The Barcelona Institute of
Science and Technology, Castelldefels (Barcelona) 08860, Spain.}

\author{Zohreh Davoudi}
\affiliation{
Maryland Center for Fundamental Physics and Department of Physics, 
University of Maryland, College Park, MD 20742, USA.}

\author{Tobias Gra\ss}
\affiliation{
ICFO-Institut de Ciencies Fotoniques, The Barcelona Institute of
Science and Technology, Castelldefels (Barcelona) 08860, Spain.}

\author{Mohammad Hafezi}
\affiliation{
Joint Quantum Institute and Department of Physics,
University of Maryland, College Park, MD 20742, USA.}
\affiliation{
Department of Electrical and Computer Engineering and \\
Institute for Research in Electronics and Applied Physics, University of Maryland, College Park, MD 20742, USA.}

\author{Guido Pagano}
\affiliation{
Department of Physics and Astronomy,
\\
Rice University, Houston, TX 77005, USA.}

\author{Alireza Seif}
\thanks{The authors' list is alphabetically ordered.}
\affiliation{
Pritzker School of Molecular Engineering
\\
University of Chicago, Chicago, Illinois 60637, USA.}

\date{\today}

\preprint{UMD-PP-021-05}

\begin{abstract}
Trapped-ion quantum simulators, in analog and digital modes, are considered a primary candidate to achieve quantum advantage in quantum simulation and quantum computation. The underlying controlled ion-laser interactions induce all-to-all two-spin interactions via the collective modes of motion through Cirac-Zoller or M{\o}lmer-S{\o}rensen schemes, leading to effective two-spin Hamiltonians, as well as two-qubit entangling gates. In this work, the M{\o}lmer-S{\o}rensen scheme is extended to induce three-spin interactions via tailored first- and second-order spin-motion couplings. The scheme enables engineering single-, two-, and three-spin interactions, and can be tuned via an enhanced protocol to simulate purely three-spin dynamics. Analytical results for the effective evolution are presented, along with detailed numerical simulations of the full dynamics to support the accuracy and feasibility of the proposed scheme for near-term applications. With a focus on quantum simulation, the advantage of a direct analog implementation of three-spin dynamics is demonstrated via the example of matter-gauge interactions in the U(1) lattice gauge theory within the quantum link model. The mapping of degrees of freedom and strategies for scaling the three-spin scheme to larger systems, are detailed, along with a discussion of the expected outcome of the simulation of the quantum link model given realistic fidelities in the upcoming experiments. The applications of the three-spin scheme go beyond the lattice gauge theory example studied here and include studies of static and dynamical phase diagrams of strongly interacting condensed-matter systems modeled by two- and three-spin Hamiltonians.
\end{abstract}

\maketitle

\section{Introduction
\label{sec:intro}}
\noindent
Quantum simulation is expected to be a promising avenue to revealing the rich dynamics of quantum many-body systems, from strongly correlated electron systems in material \cite{Mazurenko2017}, to dense nuclear matter in the interior of neutron stars~\cite{Beck:2019, cloet2019opportunities}, to coherent neutrino propagation from core-collapse supernovae and early universe~\cite{Cervia:2019res, Hall:2021rbv}, to strongly interacting gauge field theories of the Standard Model of particle physics~\cite{Preskill:2018fag, Banuls:2019bmf, Aidelsburger:2021mia, Klco:2021lap}. Quantum simulators can perform in analog, digital, or hybrid modes, with the underlying hardware architecture ranging from condensed-matter-based systems to optical, atomic, and molecular systems. As the vibrant field of quantum simulation moves toward making large-scale noise-resilient quantum simulators and quantum computers a reality, it is crucial to pursue a co-design process in which the specifications of the theoretical problem subject to the simulation, such as characteristics of interactions and degrees of freedom of the physical system, can be taken into account in enhancing and extending the simulator's toolkit. With a focus on the analog mode of operation of the quantum device, such an objective is applied in this work to trapped-ion quantum simulators~\cite{blatt2012quantum, monroe2021programmable,schneider2012experimental}. These systems have demonstrated their value in recent years in successful simulation of quantum many-body systems~\cite{islam2011onset,jurcevic2014quasiparticle,richerme2014non,zhang2017observation,hess2017non,Kaplan2020many,Pagano2020quantum,tan2021domain,kyprianidis2021observation,lanyon2011universal,Martinez:2016yna}, some of which have started to challenge classical numerical methods. These simulations were mostly based on an effective two-spin Ising Hamiltonian with single-spin field terms. Motivated by the need for an enhanced dynamics as exhibited in certain exotic spin systems as well as in lattice gauge theories, we generalize this effective Hamiltonian to a Hamiltonian with simultaneous single-, two-, and three-spin dynamics, and examine its accuracy through a numerical simulation with realistic experimental parameters.

While low-energy descriptions of interactions in nature primarily rely on two-body couplings, three- and higher-body couplings occur in many physical systems and impact the dynamics in a nontrivial manner. In high-energy physics, quantum field theories exhibit interacting terms beyond two-field couplings, including gauge-matter and gauge-field self interactions in gauge theories of the Standard Model~\cite{hokim1998, Quigg:2013, Schwartz:2014sze, Zyla:2020zbs}. In nuclear physics, three and higher-nucleon interactions that are effectively generated from quantum chromodynamics (QCD) have proven to be important in providing an accurate description of nuclei and dense matter~\cite{Kaplan:2005es,Machleidt:2011zz,Hammer:2019poc,Carlson:2014vla,Tews:2020hgp}. While present-day quantum simulators are not yet suited for simulating quantum fields and their complex interactions in general (see e.g., Refs.~\cite{Byrnes:2005qx, IgnacioCirac:2010us, Lamata:2011me, Jordan:2011ne, Jordan:2011ci, Casanova:2011wh, Tagliacozzo:2012vg, Banerjee:2012pg, Zohar:2012xf, Zohar:2013zla, hauke13, Mezzacapo:2015bra, bazavov2015gauge, Zache:2018jbt, Stryker:2018efp, Raychowdhury:2018osk, Klco:2018zqz, Bender:2018rdp, Davoudi:2019bhy, Lamm:2019bik, Harmalkar:2020mpd, Shaw:2020udc, Kharzeev:2020kgc, Chakraborty:2020uhf, Liu:2020eoa, Barata:2020jtq, Paulson:2020zjd, Ciavarella:2021nmj, Dasgupta:2020itb, gerritsma2010quantum, Martinez:2016yna, Klco:2018kyo, Lu:2018pjk, Klco:2019evd, Atas:2021ext, Kreshchuk:2020kcz, Bauer:2021gup, Gorg:2018xyc, schweizer2019floquet, Mil:2019pbt, Yang:2020yer, Davoudi:2020yln, Rahman:2021yse, Davoudi:2021ney, Stryker:2021asy, Banuls:2019bmf, Aidelsburger:2021mia, Klco:2021lap} for some recent progress), the possibility of inducing simultaneous interactions among more than two degrees of freedom of a quantum simulator can open up many interesting possibilities. A more immediate application is in the realm of condensed matter physics where it is argued that spin systems with two- and three-spin Hamiltonians exhibit novel phase diagrams and quantum phase transitions, signified by unique entanglement characteristics that are absent in systems with only two-spin couplings~\cite{pollmann2010entanglement, verresen2017one, verresen2018topology, smith2019crossing, tseng1999quantum, wannier1950antiferromagnetism, peng2009quantum, d1998level, penson1988conformal, gross1984simplest, GARDNER1985747, nieuwenhuizen1998quantum, tsomokos2008chiral, Wen2004}. Such dynamics are interesting in the context of spin glass physics~\cite{gross1984simplest, GARDNER1985747, nieuwenhuizen1998quantum} and quantum statistics~\cite{Francesco1997}. Furthermore, quantum link models that exhibit local symmetry constraints, and are argued to approximate infinite-dimensional gauge field theories in certain limits~\cite{chandrasekharan1997quantum, Brower:1997ha}, are spin systems (after mapping fermions to hard-core bosons) with multi-spin interactions. 

 Given such rich physics exhibited in systems that are governed by multi-spin interactions, and given the interest in demonstrating the near-term benefits of analog (and hybrid) quantum simulators, multitude of proposals and strategies are put forward to directly engineer correlated three-qubit interactions. These ideas include the use of Floquet engineering in superconducting qubits~\cite{liu2020synthesizing}, optimization of periodic drives in weakly driven quantum systems such as  superconducting circuits and molecular nanomagnets~\cite{petiziol2021quantum}, perturbative generation of multi-spin interactions with tunable couplings in a triangular configuration of an optical lattice of two atomic species~\cite{pachos2004three}, natural generation of three-body coupling in polar molecules driven by microwave fields~\cite{buchler2007three}, adiabatic passages
with tunneling interactions among atoms in a one-dimensional optical lattice~\cite{pachos2003quantum}, coupling Rydberg-pair interactions and collective motional modes in trapped Rydberg ion systems~\cite{gambetta2020long}, and resonantly driving three-spin interactions through an adiabatic elimination of the off-resonant transitions involving spin-phonon couplings in trapped-ion systems~\cite{bermudez2009competing}.

More specifically, the possibility of engineering a three-spin Hamiltonian in a trapped-ion quantum simulator was proposed by Bermudez et al. in Ref.~\cite{bermudez2009competing}, where the use of first and second order spin-phonon couplings enabled the generalization of the two-spin ``phase gate''\cite{Leibfried2003} Hamiltonian $\sim \sigma^z_i \sigma^z_j$ to a three-spin Hamiltonian $\sigma^z_i\sigma^z_j\sigma^z_k$, where $i,j,k$ are the ion indices. An analogous strategy is applied in this work to induce effective correlated spin-flipping transitions among (quasi)spins of three ions using nearly resonant single- and double-sideband transitions, extending upon the well-known M{\o}lmer-S{\o}rensen scheme~\cite{Molmer1999multiparticle}. Effective interactions proportional to $\sigma^+_i\sigma^-_j+{\rm h.c.}$ and $ \sigma^+_i\sigma^+_j\sigma^+_k+{\rm h.c.}$ are generated, along with a single-spin Hamiltonian proportional to $\sigma^z_i$. Off-resonant contributions constitute interactions that entangle the spin and phonon degrees of freedom and are ensured to be kept small with a careful choice of  laser and trap parameters as will be discussed. Our strategy for tuning the relative importance of the two- and three-spin dynamics is different from Ref.~\cite{bermudez2009competing} and does not require addressing the axial modes of motion. In particular, by introducing an almost ``mirror'' copy of the drives that induce effective single-, two-, and three-spin interactions, the single- and two-spin Hamiltonians can be significantly suppressed. To validate the effective dynamics, we go beyond a qualitative discussion of requisite experimental parameters and present a thorough numerical simulation of a three-spin coupling scheme in the trapped-ion system by examining the exact dynamics (within a highly accurate rotating-wave approximation), including the effects of spin-phonon contamination. High fidelities for the three-spin dynamics, as well as the effectiveness of the ``mirror'' drive in suppressing lower-body interactions, are established for realistic experimental values for laser intensities and frequencies, and for the axial trap frequency that controls the relative strength of contributions from multiple normal modes to the dynamics.

To demonstrate an application of our scheme in simplifying near-term quantum simulations of physical models of interest, the example of the lattice Schwinger model within its simplest quantum-link-model representation is examined and feasible strategies are proposed for scaling the simulations given undesired contamination from proliferation of the normal modes of motions. Furthermore, the question of the severity of local gauge-symmetry violations in light of imperfect fidelity of realistic experiments is analyzed through a crude model of interactions. The underlying scheme for Hamiltonian engineering of this work, through analog or gate-based implementations, can be extended analogously to engineer three-spin Hamiltonians with higher-dimensional spin degrees of freedom, as well as four- and higher-body spin interactions, with direct applications in lattice gauge-theory simulations. In the context of bosonic quantum field theories such as gauge theories, another interesting extension of the scheme is the engineering of effective spin-spin-phonon dynamics which will be explored in future work.

The present paper is organized as follows. Section~\ref{sec:three-spin} contains a detailed description of the scheme and its numerical simulation. In particular, Sec.~\ref{sec:schematic} offers a semi-qualitative description of the underlying mechanism for the generation of an effective three-spin Hamiltonian without full technical details. A detailed derivation of the effective Hamiltonian for single- and multi-mode scenarios and using single- and multi-drive schemes to control the three-spin dynamics is provided in Sec.~\ref{sec:derivation}. Section~\ref{sec:numerics} presents the numerical simulations of the exact unitary time evolution of the three-ion system, including phonon contributions to the dynamics, and explores the range of laser intensities and frequencies, as well as trap characteristics, that result in maximum fidelities. To present an application of the engineered three-spin Hamiltonian in the context of quantum simulation, the example of the quantum-link-model representation of  U(1) lattice gauge theory is studied in Sec.~\ref{sec:schwinger}, and includes a discussion of an exact mapping of the degrees of freedom and the Hamiltonian and the scaling specifications and limitations in Sec.~\ref{sec:QLM}, an analysis of the expected dynamics in the quantum link model assuming imperfect gauge-violating interactions in  Sec.~\ref{sec:QLM-numerics}, and a brief note on how the fully digital implementation of the model may compare with its fully analog simulation within the scheme of  this work in Sec.~\ref{sec:digital}. Conclusions and a discussion of future directions are presented in Sec.~\ref{sec:conclusion}.

\section{Engineering an effective three-spin Hamiltonian with trapped ions
\label{sec:three-spin}}
\noindent
The effective Hamiltonian that couples three spins simultaneously can be engineered using tailored ion-laser interactions. Section~\ref{sec:schematic} presents basic theoretical aspects of the ion-laser dynamics in a linear Paul trap~\cite{paul1990electromagnetic}, along with a qualitative description of an enhanced M{\o}lmer-S{\o}rensen scheme that leads to an effective three-spin Hamiltonian. Section~\ref{sec:derivation} presents the exact relations for the effective Hamiltonians, including the desired three-spin Hamiltonian and  accompanying single- and two-spin Hamiltonians, along with their detailed derivation and a description of the errors. Section~\ref{sec:numerics} presents a numerical study of the full evolution compared with the one anticipated from the effective picture in an experimentally realistic three-ion system, and identifies regimes in the tunable parameters of the scheme that are best suited for suppressing the single- and two-spin terms, giving rise to predominantly three-spin dynamics.

\subsection{Ion-laser dynamics and a qualitative description of the scheme
\label{sec:schematic}}
\begin{figure*}[t!]
\centering
\includegraphics[scale=0.65]{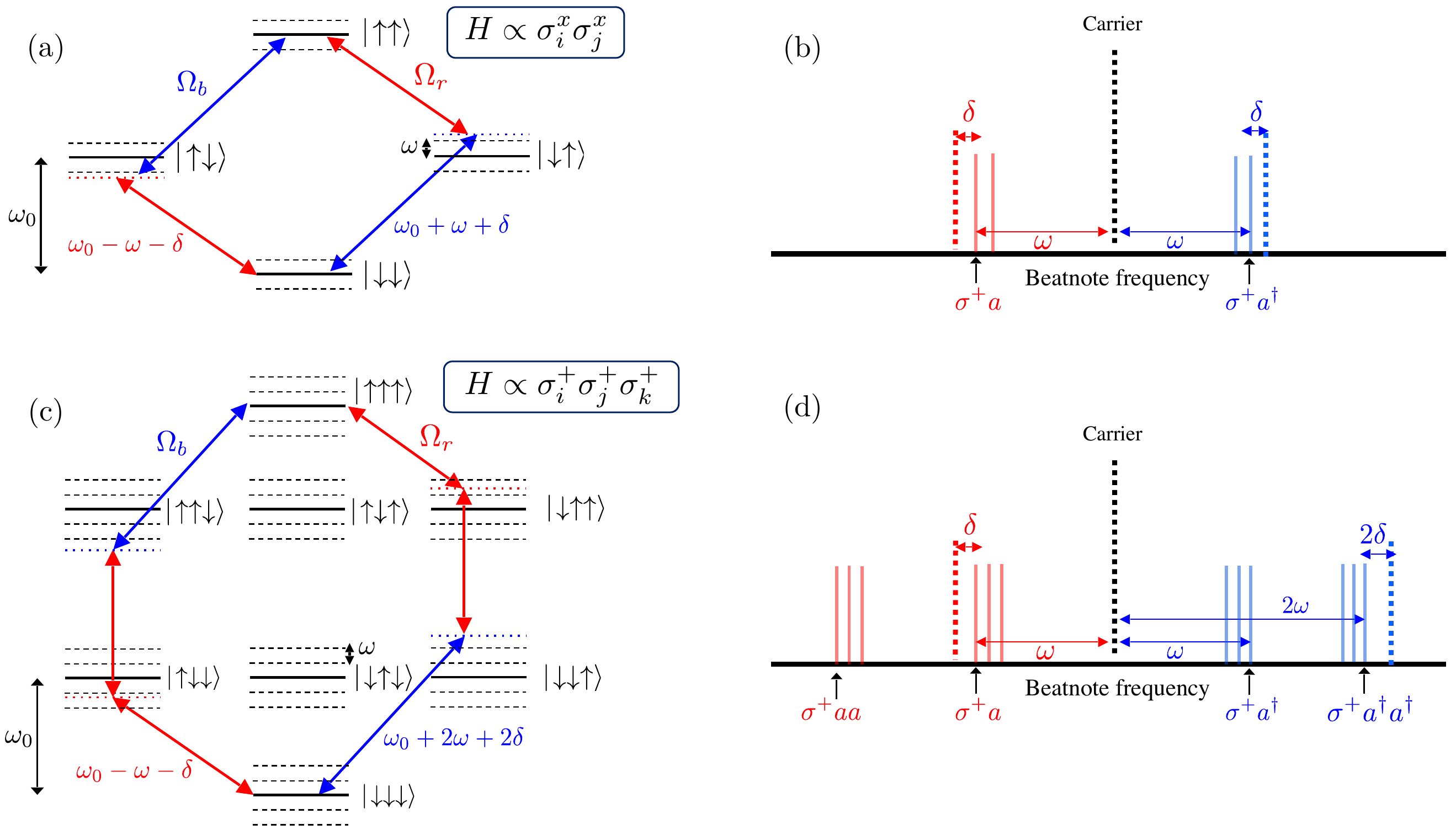}
\caption{(a,b) Traditional M{\o}lmer-S{\o}rensen scheme based on a pair of bichromatic laser beatnotes off-resonantly driving first-order spin-phonon couplings with symmetric detuning ($\pm\delta$), giving rise to an effective spin-spin interaction. The two-ion case is shown for simplicity. (c,d) Generalized M{\o}lmer-S{\o}rensen scheme to generate an effective three-spin coupling. A second-order blue sideband is driven with twice the detuning ($2\delta$) as the first-order red $(-\delta)$ sideband. As shown in (c), this process creates two virtual phonons with a second-order process and annihilates the same number of phonons through two first-order processes. Note that only two out of several possibilities are depicted. In all subfigures, $\Omega_r$ and $\Omega_b$ are the Rabi frequencies of the red and blue beatnotes, respectively. $\omega_0$ is the qubit frequency, and $\omega\,[\equiv \omega_{\rm com }]$ is the transverse center-of-mass frequency.
}
\label{fig:3vs2}
\end{figure*}

Consider a system of $N$ ions confined in a radio-frequency Paul trap. The Hamiltonian of the system in absence of any interactions with external lasers can be written as\footnote{Planck's constant $\hbar$ is set to unity throughout.}
\begin{align}
&H_{\text{free}}=\frac{\omega_0}{2}\sum_{i=1}^N \sigma_i^z+\sum_{\mathsf{m}=1}^{3N} \omega_\mathsf{m} (a_\mathsf{m}^\dagger a_\mathsf{m}+\frac{1}{2}).
\label{eq:Hfree}
\end{align}
Here, $\sigma$ is a Pauli operator acting on the space of the ions' (quasi)spin, namely the qubit. $\omega_0$ is the qubit frequency which is of the order of a few GHz for typical ions used in present-day quantum simulators. In a linear trap with a common confining potential and with long-range Coulomb interactions among the ions, the motion of the ions can be described in terms of a set collective normal modes with quantized excitations, i.e., the phonons. As a result, the displacement of each ion from its equilibrium position, $\Delta \bm{r}_i$, can be expressed in terms of phonon creation ($a_\mathsf{m}^\dagger$) and annihilation ($a_\mathsf{m}$) operators associated with the motion along the three orthogonal principal axes of the trap, $\mathsf{x},\mathsf{y}$ and $\mathsf{z}$, with commutation relations $[a_\mathsf{m},a_{\mathsf{m}'}]=[a_\mathsf{m}^\dagger,a_{\mathsf{m}'}^\dagger]=0$ and $[a_\mathsf{m},a_\mathsf{m'}^\dagger]=\delta_{\mathsf{m},\mathsf{m}'}$. Here, $\mathsf{x}$ and $\mathsf{y}$ denote the most-confined directions in the trap, which will have the same normal-mode spectra for symmetric traps commonly used. These will be denoted as transverse directions. The least-confined direction is denoted as $\mathsf{z}$ and is named the axial direction.\footnote{$\mathsf{x}$, $\mathsf{y}$ and $\mathsf{z}$ in san-serif font correspond to spatial physical coordinates of the trap, hence labeling the Cartesian components of the wave-vector of the laser beams. $x$, $y$, and $z$ in italic font correspond to the Bloch-sphere axes in the qubit Hilbert space.} In general, the indices $\mathsf{m}$ run from 1 to $3N$, but for the coupling scheme described below, only the modes along one spatial direction, selected by a light field, will be relevant. The index $\mathsf{m}$ will then run from 1 to $N$ to denote the relevant set of modes.

One or multiple pairs of counter-propagating laser beams, indexed by $L$, can be introduced with wave-vector difference $\Delta \bm{k}_L$, frequency difference (beatnote) $\Delta \omega_L$, and phase difference $\Delta \phi_L$, to stimulate the two-photon Raman transition among the two states of the qubit with a Rabi frequency $\Omega_L$. The beams can address the ions globally, or for more flexibility in devising interactions individually, in which case both amplitude and frequency control of the beams are required (and an index $i$ must be attached to laser's parameters). 

The Hamiltonian describing the ion-laser interactions can be written as~\cite{schneider2012experimental} 
\begin{eqnarray}
&&H_{\rm int.}=\sum_{i=1}^N \sum_{L=1}^{n_L} \Omega_L e^{-i\Delta\omega_L t+i\Delta\varphi_L+i\Delta \bm{k}_L \cdot \Delta \bm{r}_i}
\nonumber\\
&& \hspace{0.35 in} \times ~ (\alpha_0\mathbb{I}_i+\alpha_1\sigma^x_i+\alpha_2\sigma^y_i+\alpha_3\sigma^z_i)+\text{h.c.}
\label{eq:Hionlaser}
\end{eqnarray}
Here, index $L$ runs over $n_L$ pairs of Raman beams. $\alpha_0,\alpha_1,\alpha_2$, and $\alpha_3$ are constants related to spin-dependent forces on the two states of the qubit~\cite{schneider2012experimental} and are controlled by the intensity, geometry, and polarization of the laser beams. These are set to $\alpha_0=\alpha_2=\alpha_3=0$ and $\alpha_1=\frac{1}{2}$, associated with a common choice. The operator $\Delta \bm{k}_L \cdot \Delta \bm{r}_i$ can be written as $\sum_{\mathsf{m}=1}^{3N}\eta_{\mathsf{m},j}(a_\mathsf{m}+a_\mathsf{m}^\dagger)$, where $\eta_{\mathsf{m},i}$ are Lamb-Dicke parameters defined as $\eta_{\mathsf{m},i}=\sqrt{\tfrac{|\Delta \bm{k}|^2}{2M_{\rm ion}\omega_\mathsf{m}}}b_{\mathsf{m},i}$. Here, $b_{\mathsf{m},i}$ are the (normalized) normal-mode eigenvector components between ion
$i$ and mode $\mathsf{m}$, and $M_{\rm ion}$ denotes the mass of the ion. For optimal control, the ion-laser system must operate in the Lamb-Dicke regime where $\eta_{\mathsf{m},i}\braket{(a_\mathsf{m}+a_\mathsf{m}^\dagger)^2}^{1/2} \ll 1$. To exclusively couple the Raman beams to one set of normal-mode excitations, $\Delta \bm{k}_L$ can be set along only a single principal axis.

In the interaction picture with respect to the free Hamiltonian $H_{\rm free}$, the interacting Hamiltonian becomes
\begin{align}
&H_{\text{int.}}'=\sum_{i=1}^N \sum_{L=1}^{n_L}\frac{\Omega_L}{2} e^{i\sum_{\mathsf{m}=1}^{3N}\eta_{\mathsf{m},i}(a_\mathsf{m}e^{-i\omega_\mathsf{m} t}+a_\mathsf{m}^\dagger e^{i\omega_\mathsf{m} t})}\times
\nonumber\\
&\hspace{3.2 cm}e^{-i(\Delta \omega_L-\omega_0)t+i\Delta \phi_L} \sigma^+_i+\text{h.c}.,
\label{eq:Hintprime}
\end{align}
where it is assumed that $|\Delta \omega_L-\omega_0| \ll \omega_0$. The prime on $H$ is to denote that this Hamiltonian is in the interaction picture. Tuning the Raman-beams beatnote to $\omega_0$ ($\omega_0\pm\omega_{\mathsf{m}}$) leads to carrier (sideband) transitions, which can be derived by expanding Eq.~(\ref{eq:Hintprime}) to $\mathcal{O}(\eta^0)$ ($\mathcal{O}(\eta)$), where $\eta$ is the short-hand notation for the Lamb-Dicke parameter defined above. Similarly, an $n^{\rm th}$-order side band transition can be achieved by expanding Eq.~(\ref{eq:Hintprime}) to $\mathcal{O}(\eta^n)$ and setting $\Delta \omega_L = \omega_0\pm n\omega_{\mathsf{m}}$.

Consider one set of Raman beams. Setting the detuning $\mu \equiv \Delta \omega_L -\omega_0$ of the Raman beams to $\pm \omega_{\mathsf{m}}$, there arise interactions proportional to\footnote{Upon tunig the Raman-beam phase differences properly.}
\begin{eqnarray}
\sigma_i^{+} a_\mathsf{m},~~\sigma_i^{-} a^\dagger_\mathsf{m},~~\sigma_i^{+} a_\mathsf{m}^\dagger,~~\sigma_i^{-} a_\mathsf{m},
\label{eq:spinphonon}
 \end{eqnarray}
where $\sigma_i^\pm=\frac{1}{2}(\sigma_i^x\pm i\sigma_i^y)$. These describe spin excitation of ion $i$ accompanied by either absorption or emission of a quantum of motion in the $\mathsf{m}^{\rm th}$ mode, the so-called blue and red sideband transitions, respectively. On the other hand, when $\mu \neq \pm \omega_\mathsf{m}$ but $|\mu-\omega_\mathsf{m}|\gg\eta_{\mathsf{m},i} \Omega_L$, the motional modes are only virtually excited, and the first-order contributions in the Lamb-Dicke parameter give rise to an effective Hamiltonian proportional to
\begin{equation}
\sigma^x_i \sigma^x_j.
\label{eq:Hij}
\end{equation}
As will be shown in Sec.~\ref{sec:derivation}, in the limit where the detuning from the center-of-mass mode, $\delta$, satisfies: $|\delta| \equiv |\mu -\omega_{\rm com}| \ll \Delta_{\mathsf{m}}$ for a typical mode splitting $\Delta_\mathsf{m}$, the two-spin coupling can be approximated as $\frac{\eta_{\rm com}^2\Omega_r\Omega_b}{\delta}$, neglecting counter-rotating terms. Here, $\eta_{\rm com} = \sqrt{\tfrac{|\Delta \bm{k}|^2}{2M_{\rm ion}\omega_{\rm com}N}}$ is the Lamb-Dicke parameter associated to the center-of-mass normal mode, and $\Omega_r (\Omega_b)$ are the Rabi frequencies associated with the red and blue sideband beatnotes, respectively. This is the familiar M{\o}lmer-S{\o}rensen scheme, that is schematically described in Fig.~\ref{fig:3vs2}(a-b). In particular, it is shown how phonon emission and absorption cooperate to create a coherent coupling between the $\ket{\downarrow\downarrow}$ and $\ket{\uparrow\uparrow}$ two-ion states.

The key feature of this work, which resembles that first proposed in Ref.~\cite{bermudez2009competing}, is the generalization of the M{\o}lmer-S{\o}rensen scheme to create an effective three-spin Hamiltonian. This scheme relies on second-order contributions in the Lamb-Dicke parameter, interactions that are proportional to
\begin{eqnarray}
&&\sigma_i^+ a_\mathsf{m} a_\mathsf{n},~~\sigma_i^+ a_\mathsf{m} a^\dagger_\mathsf{n},~~\sigma_i^+ a_\mathsf{m}^\dagger a_\mathsf{n},~~\sigma_i^+ a_\mathsf{m} a_\mathsf{n},
\nonumber\\
&&\sigma_i^- a^\dagger_\mathsf{m} a^\dagger_\mathsf{n},~~\sigma_i^- a^\dagger_\mathsf{m} a_\mathsf{n},~~\sigma_i^- a_\mathsf{m} a^\dagger_\mathsf{n},~~\sigma_i^- a^\dagger_\mathsf{m} a^\dagger_\mathsf{n}.
\end{eqnarray}
Here, a spin excitation can be associated with the creation or annihilation of two phonons at the same time. As shown in Fig.~\ref{fig:3vs2}(c-d), this scheme amounts to driving the ion chain using two beatnotes with asymmetric detunings (for example, $2\delta$ for the blue and $-\delta$ for the red sideband) so that a single second-order spin-phonon excitation process can be combined with two first-order absorption processes, giving rise to a coherent resonant coupling between $\ket{\downarrow\downarrow\downarrow}$ and $\ket{\uparrow\uparrow\uparrow}$ states. This leads to an effective three-spin Hamiltonian proportional to
\begin{equation}
\label{eq_H3}
\sigma_i^+ \sigma_j^+ \sigma_k^+,~~\sigma_j^- \sigma_j^- \sigma_k^-,
\end{equation}
with a coupling that can be approximated by $\frac{\eta_{\rm com}^4 \Omega_r^2\Omega_b}{\delta^2}$, in the limit of small detuning $\delta$ from the center-of-mass mode. This process generates resonant single- and two-spin interactions as well, and if the goal is to achieve a pure three-spin Hamiltonian, these contributions must be suppressed, particularly given that their strength is larger than that of the three-spin interactions. For this reason, a second drive is added to cancel out the undesired single- and two-spin contributions that are proportional to $\frac{1}{\delta}$ and will enhance the three-spin contribution that is proportional to $\frac{1}{\delta^2}$. This additional drive is composed of two sets of Raman beams with asymmetric detunings of opposite sign compared with the first drive i.e., $\sim -2\delta$ for the blue and $\sim \delta$ for the red sideband. This cancellation is only exact for the center-of-mass mode contribution. The exact relations in the next section will provide further essential detail of the single- and two-drive schemes in the single- and multi-mode scenarios, and the numerical study of Sec.~\ref{sec:numerics} will investigate the accuracy of the effective three-spin dynamics.

\subsection{Derivation of the effective Hamiltonian
\label{sec:derivation}}
To obtain the exact relations for the effective dynamics of the ion-laser system within the scheme just described, one can start from the interaction-picture Hamiltonian in Eq.~(\ref{eq:Hintprime}). To engineer an effective three-spin Hamiltonian, we introduce two pairs of Raman beams, as described in the previous section. The Rabi and beatnote frequencies associated with pair $I$ are denoted as $\Omega_r$ and $\mu_r \equiv \Delta \omega_0-\omega_I$, respectively, and those associated with pair $II$ are denoted as $\Omega_b$ and $\mu_b \equiv \omega_0-\Delta \omega_{II}$. Additionally, the following relations
\begin{align}
&\mu_r=-\omega_{\rm com}-\delta,
\label{eq:mur}
\\
&\mu_b=-2\mu_r=2\omega_{\rm com}+2\delta,
\label{eq:mub}
\end{align}
will be applied to ensure resonant three-spin transitions, as will be seen shortly. This choice of beatnote frequencies justifies the use of `$r$' and `$b$' subscripts for each pair, corresponding to (single) red and (double) blue sideband transitions of the qubit, respectively. With this setup, and ignoring contributions from counter-rotating waves oscillating with a frequency of $\omega_\mathsf{m}$ or higher, the Hamiltonian of the system at $\mathcal{O}(\eta)$ in the interaction picture becomes
\begin{align}
&H_{\rm int.}^{\rm RWA} \big |_{\mathcal{O}(\eta)}=\frac{i}{2}\sum_{i=1}^N\sum_{\mathsf{m}=1}^N \eta_{\mathsf{m},i}
\Omega_r e^{i(\omega_{\mathsf{m}}+\mu_r)t}a^\dagger_{\mathsf{m}}\sigma^+_i+{\rm h.c.},
\label{eq:Heta}
\end{align}
and that at $\mathcal{O}(\eta^2)$ reads
\begin{align}
&H_{\rm int.}^{\rm RWA} \big |_{\mathcal{O}(\eta^2)}=-\frac{1}{4}\sum_{i=1}^N\sum_{\mathsf{m}=1}^N\sum_{\mathsf{n}=1}^N \eta_{\mathsf{m},i} \eta_{\mathsf{n},i} \Omega_b e^{-i(\omega_{\mathsf{m}}+\omega_{\mathsf{n}}-\mu_b)t}
\nonumber\\
&\hspace{5.35 cm} a_{\mathsf{m}}a_{\mathsf{n}}\sigma^+_i+{\rm h.c.},
\label{eq:Hetaeta}
\end{align}
where RWA in the superscripts refers to the rotating-wave approximation applied. Only the transverse normal modes along the $\mathsf{x}$ direction of the trap are addressed with both pairs of the beams, hence the sums over modes $\mathsf{m}$ and $\mathsf{n}$ run from $1$ to $N$.

Performing a Magnus expansion~\cite{Magnus:1954zz} of the time-evolution operator, keeping only contributions that are resonant and further ignoring contributions from counter-rotating terms (see discussions below), give rise to an effective (time-independent) spin Hamiltonian. Explicitly,
\begin{eqnarray}
U(t,0)&=&\mathcal{T} e^{-i\int_0^t \big(H_{\rm int.}'(\tau) \big |_{\mathcal{O}(\eta)}+H_{\rm int.}'(\tau) \big |_{\mathcal{O}(\eta^2)}\big)d\tau}
\nonumber\\
&\approx& e^{-i \big(H^{(\sigma)}_{\rm eff}+H^{(\sigma\sigma)}_{\rm eff}+H^{(\sigma\sigma\sigma)}_{\rm eff}\big) t},
\label{eq:U}
\end{eqnarray}
where $\mathcal{T}$ denotes time ordering. The single-spin Hamiltonian has the form
\begin{align}
&H^{(\sigma)}_{\rm eff} = \frac{1}{4}\sum_{i} \bigg\{ \sum_{\mathsf{m}} \eta_{\mathsf{m},i}^2\frac{\Omega_r^2}{\omega_{\mathsf{m}}+\mu_r}(n_{\mathsf{m}}+\frac{1}{2})
\nonumber\\
& \hspace{0.8 cm} -\frac{1}{4}\sum_{\mathsf{m},\mathsf{n}}  \frac{\Omega_b^2}{\omega_{\mathsf{m}}+\omega_{\mathsf{n}}+2\mu_r}\,\bigg[\eta_{\mathsf{m},i}^2 \eta_{\mathsf{n},i}^2(n_{\mathsf{m}}+n_{\mathsf{n}}+1)\, +
\nonumber\\
& \hspace{0.65 cm} \sum_{\mathsf{m}',\mathsf{n}'}\eta_{\mathsf{m},i} \eta_{\mathsf{n},j}\eta_{\mathsf{m'},i} \eta_{\mathsf{n'},j} \delta_{\omega_{\mathsf{m}'}+\omega_{\mathsf{n}'},\omega_{\mathsf{m}}+\omega_{\mathsf{n}}}\, a^\dagger_{\mathsf{m}'}a^\dagger_{\mathsf{n}'}a_{\mathsf{m}}a_{\mathsf{n}}\bigg] \bigg\} \, \sigma^z_i.
\label{eq:H1sigmamulti}
\end{align}
The two-spin Hamiltonian reads
\begin{align}
&H^{(\sigma \sigma)}_{\rm eff} = -\frac{1}{4}\sum_{i} \sum_{i \neq j} \bigg\{ \sum_{\mathsf{m}} \eta_{\mathsf{m},i}\eta_{\mathsf{m},j}\frac{\Omega_r^2}{\omega_{\mathsf{m}}+\mu_r}+\frac{1}{2}\sum_{\mathsf{m},\mathsf{n}}
\nonumber\\
& \hspace{0.35 cm} \eta_{\mathsf{m},i}\eta_{\mathsf{n},i} \eta_{\mathsf{m},j}\eta_{\mathsf{n},j} \frac{\Omega_b^2}{\omega_{\mathsf{m}}+\omega_{\mathsf{n}}+2\mu_r} (n_{\mathsf{m}}+n_{\mathsf{n}}+1)\bigg\}\sigma^+_i\sigma^-_j,
\label{eq:H2sigmamulti}
\end{align}
and the desired three-spin Hamiltonian takes the form
\begin{align}
&H^{(\sigma \sigma \sigma)}_{\rm eff} = \sum_{i,j,k} \sum_{\mathsf{m},\mathsf{n}} \eta_{\mathsf{m},i}\eta_{\mathsf{n},j} \eta_{\mathsf{m},k}\eta_{\mathsf{n},k} \, \Omega_r^2\Omega_b \, \sigma^+_i\sigma^+_j\sigma^+_k
\nonumber\\
&\hspace{1.5 cm}\frac{3\mu_r+\omega_\mathsf{m}+2\omega_\mathsf{n}}{24(\mu_r+\omega_\mathsf{m})(\mu_r+\omega_\mathsf{n})(2\mu_r+\omega_\mathsf{m}+\omega_\mathsf{n})}.
\label{eq:H3sigmamulti}
\end{align}
In these relations, $n_{\mathsf{m}(\mathsf{n})} \equiv a^\dagger_{\mathsf{m}(\mathsf{n})}a_{\mathsf{m}(\mathsf{n})}$ is the occupation-number operator for the corresponding phonon mode, and the condition $\mu_b=-2\mu_r$ is imposed. In an individual-addressing scheme, the Rabi frequencies can be set independently at the location of each ion, hence the following replacements must be taken into account: $\Omega_{r(b)}^2 \to \Omega_{r(b),i}^2$ in Eq.~(\ref{eq:H1sigmamulti}), $\Omega_{r(b)}^2 \to \Omega_{r(b),i}\Omega_{r(b),j}$ in Eq.~(\ref{eq:H2sigmamulti}), and $\Omega_r^2\Omega_b \to \Omega_{r,i}\Omega_{r,j}\Omega_{b,k}$ in Eq.~(\ref{eq:H3sigmamulti}).

Corrections to the above picture can be attributed to at least three sources. First, off-resonant contributions, i.e., those introducing oscillatory time dependence in the exponent of $U(t,0)$ (instead of linear time dependence that is a feature of resonant contributions), introduce corrections to the effective Hamiltonians. They include first-order terms in the Lamb-Dicke parameter and can have a time-dependence as slow as $e^{i\delta t}$. At long times and as long as $\eta \Omega_{r(b)}/\delta$ remains small, these contributions can be ignored. Such a slow time dependence occurs at higher orders in the Lamb-Dicke parameter as well, but they will average to zero compared with the resonant contributions at long times, i.e., when $t \gg 1/\delta$. Second, there are higher-order resonant and off-resonant terms that are suppressed by positive powers of Lamb-Dicke parameter compared with the contributions retained. The third source of corrections to the effective Hamiltonians are from counter-rotating terms. While being resonant, these generally scale with inverse powers of mode frequency or mode-frequency separation (and combinations of), which are suppressed compared with those included in Eqs.~(\ref{eq:H1sigmamulti})-(\ref{eq:H3sigmamulti}), which scale as inverse powers of $\delta$, provided that $\delta$ is the smallest frequency scale in the setting. Given these sources of corrections to the effective picture above, it is important to benchmark the scheme in a complete numerical simulation, and obtain the range of experimental parameters with which the effective dynamics above can be achieved. Such an investigation is conducted and described in detail in the next section.

While the simultaneous one-, two-, and three-spin dynamics may be relevant in describing a generic quantum many-body system, the overarching goal of our study is to achieve a pure three-spin Hamiltonian for applications in quantum computing, and in quantum simulation of certain lattice gauge theories. The single- and two-spin dynamics arising from Eqs.~(\ref{eq:H1sigmamulti}) and (\ref{eq:H2sigmamulti}) can in general be significant compared with the three-spin dynamics from Eq.~(\ref{eq:H3sigmamulti}). Nonetheless, there is a simplifying limit in which a scheme can be devised to cancel the undesired single- and two-spin contributions entirely. Before introducing this scheme, let us consider the following simplification. If the spacing among the modes is much larger than $\delta$, the only mode contributing to the effective dynamics is that close to the lasers beatnote frequency. This can be achieved for the transverse modes if trap's axial confinement is increased. In the scheme adopted above, this is the center-of-mass mode, see Eqs.~(\ref{eq:mur}) and (\ref{eq:mub}). Ignoring contributions from all other modes, the effective single-, two-, and three-spin Hamiltonians become
\begin{align}
&H^{(\sigma)}_{\rm eff} = -\frac{1}{4}\sum_{i} \frac{\eta_{\rm com}^2}{\delta} \bigg\{ \Omega_r^2 \, (n_{\rm com}+\frac{1}{2})
\nonumber\\
& \hspace{1.85 cm} -\frac{1}{8}\, \eta_{\rm com}^2\Omega_b^2\,(n_{\rm com}^2+n_{\rm com}+1)\bigg\} \sigma^z_i,
\label{eq:H1sigmasingle}
\end{align}
\begin{align}
&H^{(\sigma \sigma)}_{\rm eff} = \frac{1}{4}\sum_{i}\sum_{j\neq i} \frac{\eta_{\rm com}^2}{\delta} \bigg\{ \Omega_r^2 \, +
\nonumber\\
& \hspace{2.0 cm} \frac{1}{2}\,\eta_{\rm com}^2\Omega_b^2\,(n_{\rm com}+\frac{1}{2}) \bigg\}\sigma^+_i\sigma^-_j,
\label{eq:H2sigmasingle}
\end{align}
\begin{align}
&H^{(\sigma \sigma \sigma)}_{\rm eff} = \sum_{i,j,k} \frac{\eta_{\rm com}^4}{16 \, \delta^2}\Omega_r^2 \Omega_b \, \sigma^+_i\sigma^+_j\sigma^+_k+{\rm h.c.},
\label{eq:H3sigmasingle}
\end{align}
respectively. $n_{\rm com}$ is the occupation-number operator for the center-of-mass mode, i.e., $n_{\rm com} \equiv a^\dagger_{\rm com}a_{\rm com}$.\footnote{Note that for the center of mass mode, $b_{{\rm com},i}=\frac{1}{\sqrt{N}}\{1,1,\cdots,1\}$, hence $\eta_{{\rm com},i}$ can be replaced with $\eta_{\rm com}$ for each $i$, where $\eta_{\rm com}$ is defined after Eq.~(\ref{eq:Hij})}

The above single-mode approximation makes it clear that the single- and two-spin couplings are odd under $\delta \to -\delta$ while the three-spin coupling is even. Therefore, the net single- and two-spin contributions can be canceled, by introducing two extra Raman-beam pairs ($I'$ and $II'$) each with Rabi and beatnote frequencies $\Omega_r'=\Omega_r$, $\mu_r'=-\omega_{\rm com}+\delta$, and $\Omega_b'=\Omega_b$, $\mu_b'=2\omega_{\rm com}-2\delta$, respectively.  However, such detunings give rise to resonant sideband transitions when applied along with the first set of lasers. To avoid such undesired contributions and still achieve the desired cancellation, one can set the detuning of the second set of lasers from the center-of-mass mode, $\delta'$, to be incommensurate with respect to that of the first set, $\delta$. The cancellation of single- and two-spin couplings can be still achieved by scaling the Rabi frequencies accordingly. Explicitly, assuming $\delta' = q \delta$ for non-rational constant $q$
\begin{align}
&\Omega_r'=\sqrt{q}\Omega_r,~\mu_r'=-\omega_{\rm com}+q \delta,
\nonumber
\\
&\Omega_b'=\sqrt{q}\Omega_b,~\mu_b'=2\omega_{\rm com}-2 q \delta,
\label{eq:primeparam}
\end{align}
Once again, given the approximation made, it is important to investigate how close the simplified dynamics is to the full dynamics, as will be studied in the next section. The original scheme introduced earlier will be denoted in the following as a ``single-drive" scheme while the enhanced scenario is denoted as a ``two-drive" scheme.

\subsection{Numerical simulations}
\label{sec:numerics}
The effective Hamiltonian obtained from the Magnus expansion relies on the assumption that only the resonant combination of second- and third-order processes contribute to the system dynamics. In the present section, we validate and corroborate the analytical results of Sec.~\ref{sec:derivation} by performing a numerical simulation of the system dynamics. 

The numerical treatment requires discretization of time, with time steps smaller than the time scale of the fastest processes in the Hamiltonian. Therefore, to facilitate the simulation, the rotating-wave approximation in Eqs.~(\ref{eq:Heta})-(\ref{eq:Hetaeta}) is assumed, which is justified by the fact that $|\delta| \equiv |\mu-\omega_{\rm com}| \ll \omega_{\mathsf{m}}$. The dynamics generated by these Hamiltonians can differ from the desired effective spin dynamics due to off-resonant terms which are neglected in the effective picture. These terms manifest themselves by changes in the phonon number during the time evolution. Moreover, as explained above, the effective dynamics becomes a pure three-spin interaction when a second drive cancels the undesired resonant single- and two-spin contributions in Eqs.~(\ref{eq:H1sigmasingle})-(\ref{eq:H2sigmasingle}). However, this cancellation scheme is complete only with respect to the contributions from the center-of-mass mode. Therefore, the presence of more than one mode yields another source of error if the purpose is the design of a pure three-spin Hamiltonian.

\subsubsection{Single-mode approximation}
In order to discern between these two sources of errors, one can first study the effect of off-resonant terms in the single-mode approximation, where the contributions from all modes except for the center-of-mass mode are discarded. Up to second order in $\eta$, the Hamiltonian of one Raman pair reads
\begin{align}
H_{\rm 1drive}^{\rm com}=&\:\sum_{i=1}^N \Big( \frac{i\Omega_r}{2} \eta_{\rm com} e^{-i\delta t}a^\dagger_{\rm com}\sigma^+_i \nonumber \\
&\hspace{1 cm} -\frac{\Omega_b}{4} \eta_{\rm com}^2 e^{2i\delta t} \, a_{\rm com}^2\sigma^+_i \Big) +{\rm h.c.}
\label{eq:H1pair}
\end{align}
A second Raman pair with opposite detuning would cancel the undesired single- and two-spin terms from the first pair, but it would also produce a new resonant second-order term from the combination of the two pairs. Therefore, as introduced in Eq.~(\ref{eq:primeparam}), an asymmetry between the two pairs needs to be introduced through an appropriate scale factor $q\neq 1$. With this, the Hamiltonian including both drives is given by
\begin{align}
 &H_{\rm 2drive}^{\rm com}=\:\sum_{i=1}^N\bigg[ \frac{i\Omega_r}{2}\eta_{\rm com}\Big(
 e^{-i\delta t}+ \sqrt{q} e^{iq\delta t} \Big)
 a^\dagger_{\rm com}\sigma^+_i \nonumber\\
 & \hspace{0.5 cm} -\frac{\Omega_b}{4} \eta_{\rm com}^2 \Big( e^{2i\delta t} + \sqrt{q} e^{-2iq\delta t}\Big) a_{\rm com}^2\sigma^+_i\bigg]+{\rm h.c.}
\label{eq:H2pair}
\end{align}
\begin{figure*}[t!]
\centering
\includegraphics[width=1.5\columnwidth]{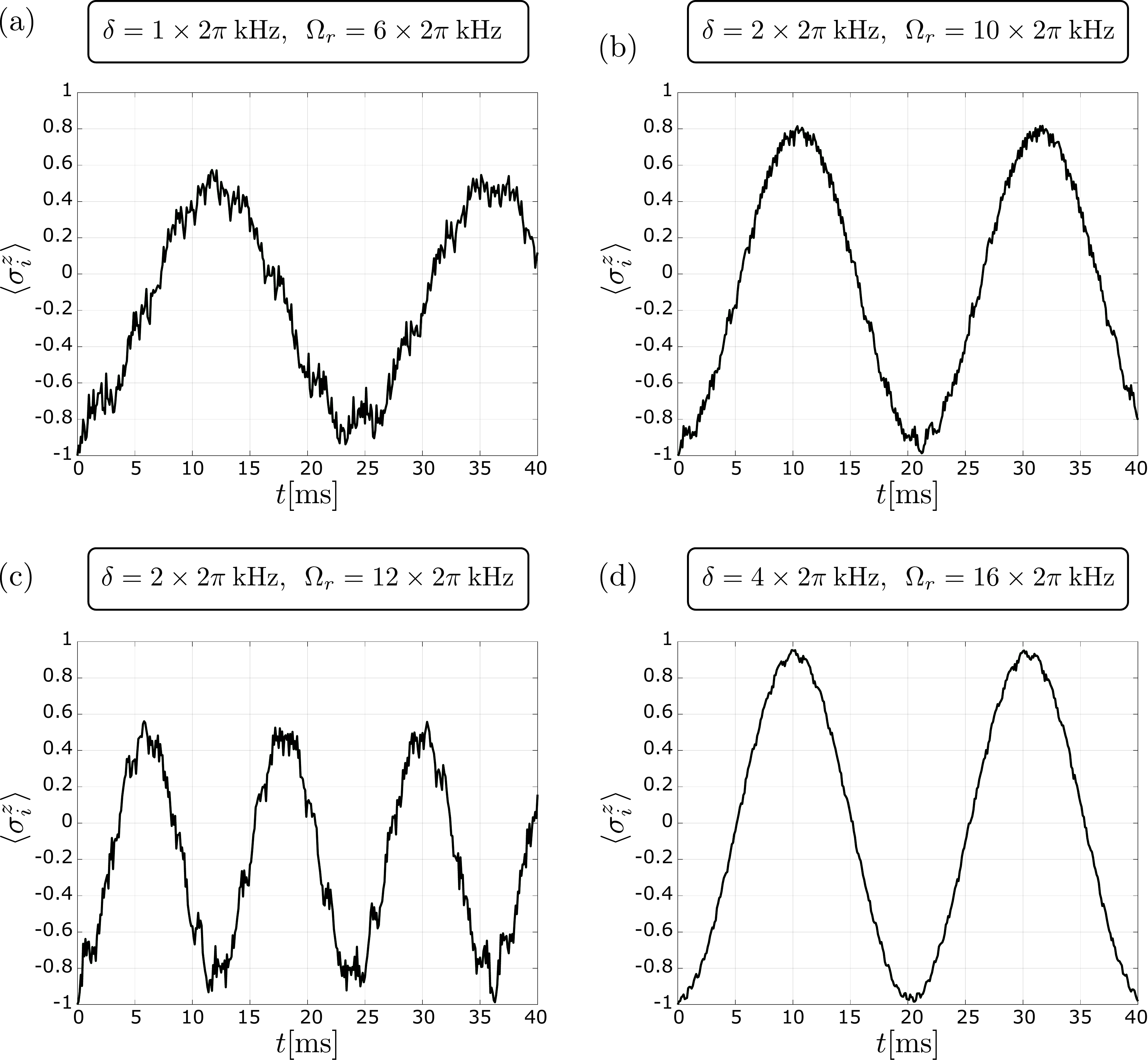}
\caption{Evolution of $\expval{\sigma_i^z}$ under the single-mode two-drive Hamiltonian $H_{\rm 2drive}^{\rm com}$ in Eq.~(\ref{eq:H2pair}), where $i=1,2,3$ is the ion index, for different choices of detuning $\delta$ and Rabi frequency $\Omega_r$. The evolution of the spins from an initial state $\ket{\downarrow \downarrow \downarrow}\otimes\ket{0}$ probes the system's three-spin interactions, and it is shown by the black curve (for ion 1), as well as fully hidden (identical) curves for ions 2 and 3.
}
\label{fig:com3body}
\end{figure*}
\begin{figure*}[t!]
\centering
\includegraphics[width=1.5\columnwidth]{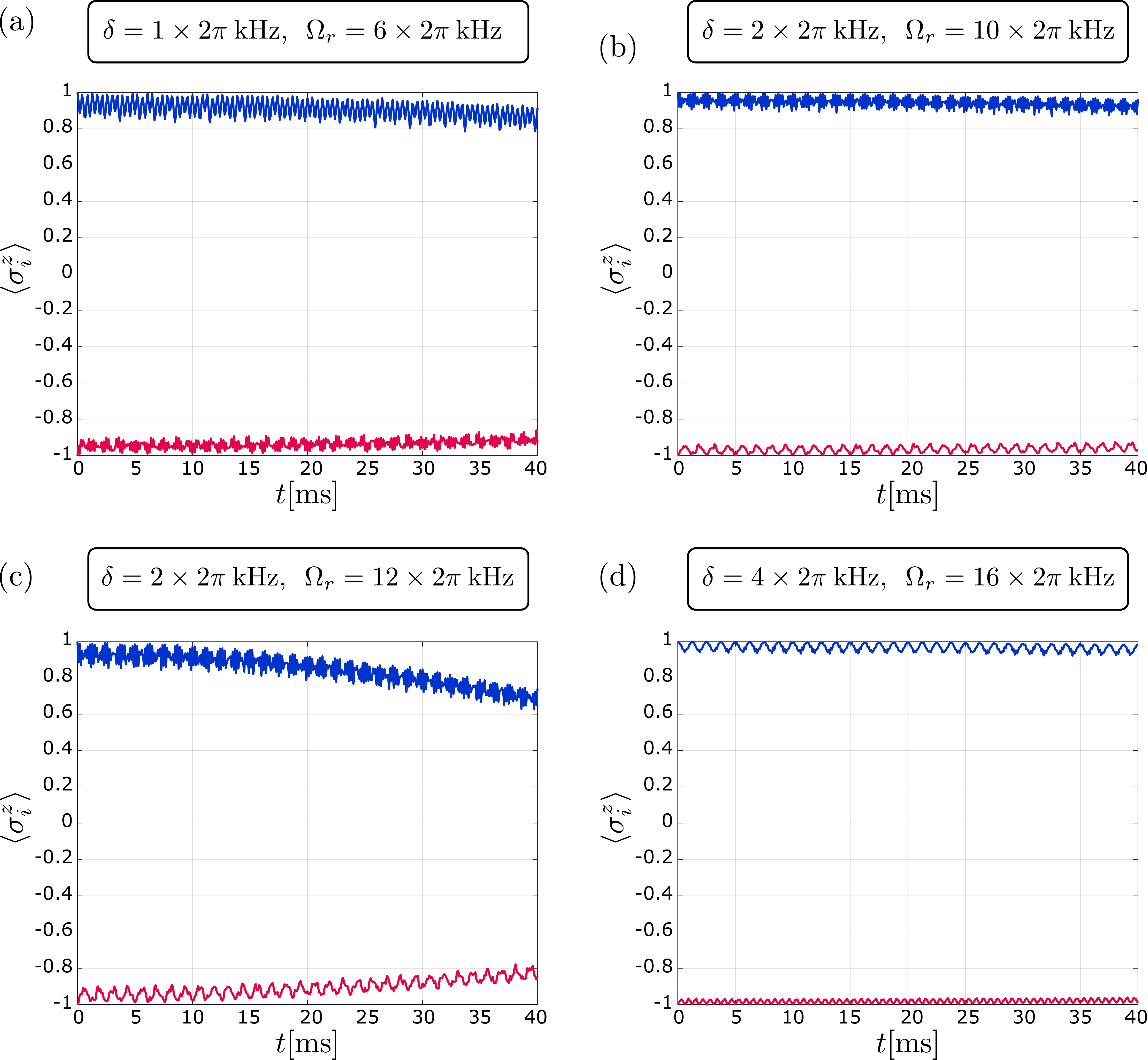}
\caption{Evolution of $\expval{\sigma_i^z}$ under the single-mode two-drive Hamiltonian $H_{\rm 2drive}^{\rm com}$ in Eq.~(\ref{eq:H2pair}), where $i=1,2,3$ is the ion index, for different choices of detuning $\delta$ and Rabi frequency $\Omega_r$. The evolution of the spins from an initial state $\ket{\downarrow \uparrow \downarrow}\otimes\ket{0}$ probes the system's two-body dynamics, and it is shown by the yellow, blue and red curves for ions 1, 2 and 3, respectively. Here, the yellow curve is fully hidden by the red one.
\color{black}
}
\label{fig:com2body}
\end{figure*}
For convenience, we choose $\Omega_b=2\Omega_r/\eta_{\rm com}$, such that the red- and blue-sideband processes appear with equal strengths. Additionally, the scale factor is set to $q=1.3$, which as will be evident shortly, appears to be a good choice. Assuming a transverse trap frequency $\omega_{\mathsf{x}}=5 \times 2\pi$~MHz, and a recoil energy of $\omega_{\rm rec}\equiv |\Delta \bm{k}|^2/(2M_{\rm ion})=26 \times 2\pi$~kHz, the Lamb-Dicke parameter for $N=3$ ions evaluates to $\eta_{\rm com}= \sqrt{\omega_{\rm rec}/(N\omega_{\mathsf{x}})} \approx 0.0416$.\footnote{Values that are accessible in current experimental trapped-ion systems operating using Ytterbium ions in radio-frequency Paul traps.} For different choices of detuning $\delta$ and Rabi frequency $\Omega_r$, the dynamics are simulated under the two-drive single-mode Hamiltonian in Eq.~(\ref{eq:H2pair}), with the results shown in Figs.~\ref{fig:com3body} and ~\ref{fig:com2body}.

Specifically, Fig.~\ref{fig:com3body} shows in black the evolution of the spin expectation value $\langle \sigma_i^z \rangle$ as a function of time if the system is initially prepared in a state $\ket{\downarrow \downarrow \downarrow}\otimes\ket{0}$ corresponding to all spins pointing down and the phonon vacuum. In all cases, this evolution exhibits an oscillatory behavior which is a manifestation of the three-spin interactions. In addition, blue and red lines in Fig.~\ref{fig:com2body} show the evolution if the system is initially prepared in the $\ket{\uparrow \downarrow \downarrow} \otimes \ket{0}$ state. In this case, the dynamics is strongly suppressed, demonstrating that the cancellation of two-body terms indeed works very well.

\begin{figure}[t!]
\centering
\includegraphics[width=0.75\columnwidth]{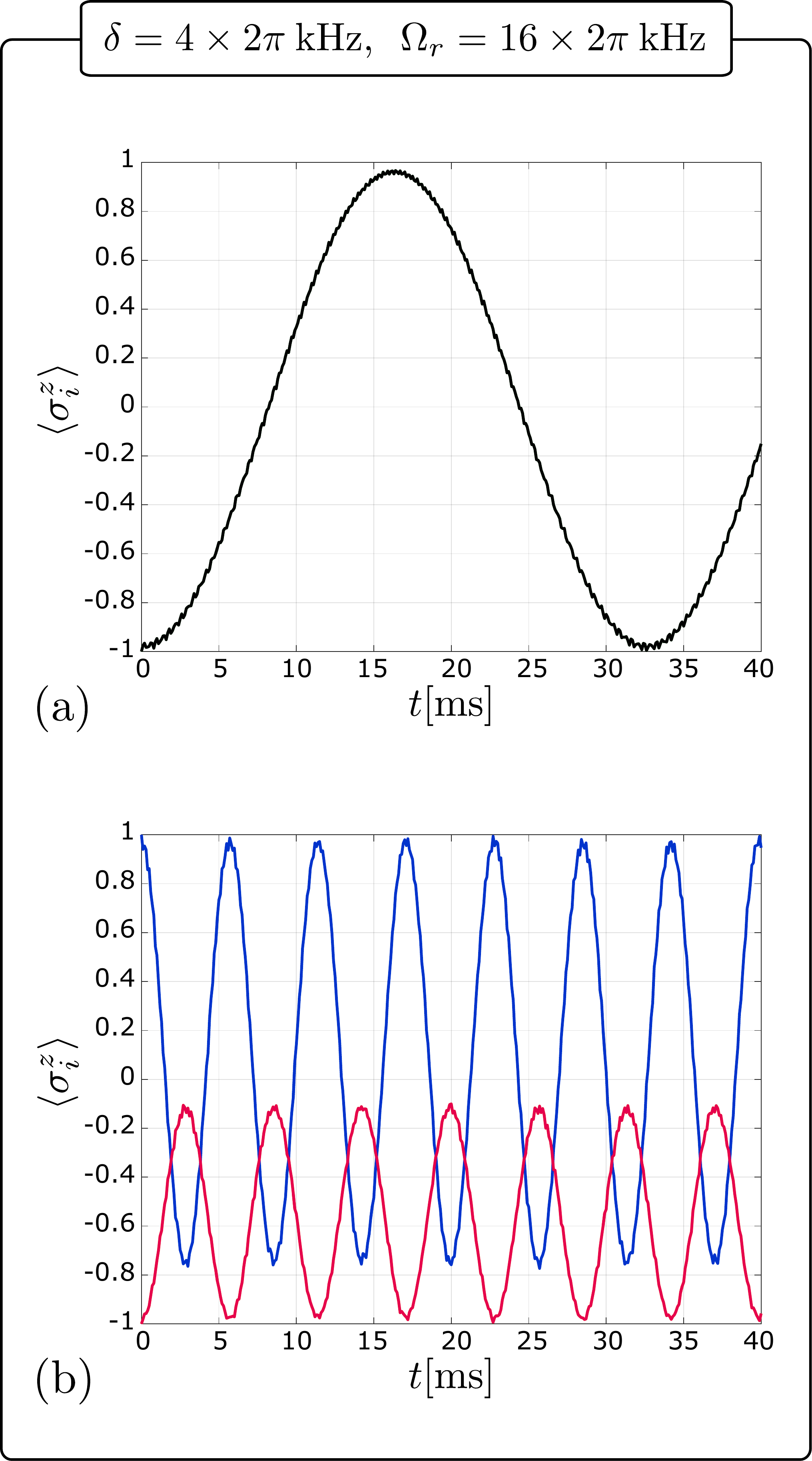}
\caption{Evolution of $\expval{\sigma_i^z}$ under a single-mode (center-of-mass) Hamiltonian with a single drive corresponding to $H_{\rm 1drive}^{\rm com}$ in Eq.~(\ref{eq:H1pair}), where $i=1,2,3$ is the ion index, for detuning and Rabi frequency shown. (a) The evolution from an initial state $\ket{\downarrow \downarrow \downarrow}\otimes\ket{0}$ probes three-spin interactions. They give rise to three-spin oscillations, represented by the black curve for ion 1. Ions 2 and 3 behave identically. (b) The two-spin dynamics is probed by considering the evolution from an initial state $\ket{\downarrow \uparrow \downarrow}\otimes\ket{0}$. Yellow, blue, and red curves correspond to ions $1,2,3$, respectively. Note that the yellow curve is fully hidden by the red one. These plots should be compared with Figs.~\ref{fig:com3body}(d) and ~\ref{fig:com2body}(d), that is the evolution under two-drive scenario using the same laser parameters. }
\label{fig:singlevsdouble_drives}
\end{figure}

To compare more quantitatively the three-spin dynamics in Fig.~\ref{fig:com3body} with the prediction from Eq.~(\ref{eq:H3sigmasingle}), one can consider the period of the oscillations in the full and the effective dynamics. In the effective Hamiltonian, the three-spin coupling from a single drive is $J_3^{\rm 1drive} \equiv \frac{\eta_{\rm com}^4 \Omega_r^2 \Omega_b}{16 \delta^2}$, and therefore, from the combination of the two drives, one expects three-spin Rabi oscillations with a period
\begin{align}
 T_3 = \frac{\pi}{6 J_3^{\rm 2drive}},
\end{align}
where $J_3^{\rm 2drive} \equiv J_3^{\rm 1 drive}(1+\sqrt{q^{-1}})$ and the factor 6 in the denominator accounts for all permutations of the three spins.
For the parameters specified in Fig.~\ref{fig:com3body}(a-d), the expected oscillations periods are: (a) $T_3 \approx 22.8$ ms, (b) $T_3 \approx 19.7$ ms, (c) $T_3 \approx 11.4$ ms, (d) $T_3 \approx 19.2$ ms. These values agree well with the observed period, showing for all cases the presence of three-spin interactions with approximately the strength predicted by Eq.~(\ref{eq:H3sigmasingle}). Another feature of the evolution plots is that, qualitatively, the oscillations are quite different in the four scenarios shown in Fig.~\ref{fig:com3body}. Clean sinusoidal oscillations, with an amplitude ranging from -1 to 1, are seen in panel (d), which among all shown cases has the largest detuning ($\delta=4 \times 2\pi$~kHz). In this case, the suppression of two-body terms, as shown in Fig.~\ref{fig:com2body}(d), also works best. 
In panels (a) and (b) of Fig.~\ref{fig:com3body}, similarly fast three-spin dynamics as in panel (d) are achieved, but at smaller values of the detuning. It is seen that for $\delta=2 \times 2\pi$ kHz in panel (b) the results are still in good accordance with sine-like oscillations, but the fidelity of the evolution becomes significantly poorer at $\delta=1 \times 2\pi$~kHz in panel (a). Similarly, the fidelity decreases if the oscillation are sped up by increasing $\Omega_r$. In this respect, one can compare panels (b) and (c), that exhibit identical detuning, but different Rabi frequencies. Clearly, the price for the speed-up in (c) is a significant loss of fidelity, also resulting in an increase of spurious two-body interactions, as shown in Fig.~\ref{fig:com2body}(c). Specifically, in Fig.~\ref{fig:com2body} we demonstrate the two-spin dynamics by initializing the system in a mixed-spin state $\ket{\downarrow\uparrow\downarrow}\otimes\ket{0}$. As is seen in Fig.~\ref{fig:com2body}(d) the maximum suppression of two-spin dynamics occurs for the largest detuning employed, which is consistent with the highest fidelity of three-spin oscillations in Fig.~\ref{fig:com3body}(d).

For a direct comparison between single- and two-drive schemes, in Fig.~\ref{fig:singlevsdouble_drives} we have simulated the dynamics of the Hamiltonian with only one Raman pair, Eq.~(\ref{eq:H1pair}). As expected, when the system is initialized in a state with all-parallel spins, the single-drive scheme in Fig.~\ref{fig:singlevsdouble_drives}(a) leads to similar three-spin oscillations as the two-drive scheme in Fig.~\ref{fig:com3body}(d), yet with approximately twice the period. However, when the initial state contains anti-parallel spins, as considered in Fig.~\ref{fig:singlevsdouble_drives}(b), the single-drive Hamiltonian exhibits very fast two-spin dynamics which is absent in the two-drive scheme in Fig.~\ref{fig:com2body}(d).

This interpretation of the spin dynamics is corroborated by the number of phonons generated during the evolution, as a quantitative measure for the deviation from the ideal evolution in which the systems remains in the phonon vacuum throughout. The average number of phonons over a time duration of 40 ms is given in Tables~\ref{tab:phononcomI} and \ref{tab:phononcomII} for the scenarios shown in Figs.~\ref{fig:com3body}, ~\ref{fig:com2body} and \ref{fig:singlevsdouble_drives}. Smaller detunings and/or larger Rabi frequencies lead to larger deviations from the phonon vacuum. The number of phonons oscillates around this time-averaged value but its peak remains below one at all times.
\begin{table}[h!]
\centering
\begin{tabular}{cccccccccccc}
\hline 
 &  &  &  &  &  &  &  &  &  &  & \tabularnewline
 & $\frac{1}{2\pi}\{\delta,\Omega_{r}\}$[kHz] &  &  & $\{1,6\}$ &  & $\{2,10\}$ &  & $\{2,12\}$ &  & $\{4,16\}$ & \tabularnewline
 &  &  &  &  &  &  &  &  &  &  & \tabularnewline
\hline 
 &  &  &  &  &  &  &  &  &  &  & \tabularnewline
 & $\left|\downarrow\downarrow\downarrow\right\rangle \otimes\left|0\right\rangle $ &  &  & 0.43 &  & 0.19 &  & 0.44 &  & 0.08 & \tabularnewline
 &  &  &  &  &  &  &  &  &  &  & \tabularnewline
 & $\left|\downarrow\uparrow\downarrow\right\rangle \otimes\left|0\right\rangle $ &  &  & 0.14 &  & 0.08 &  & 0.14 &  & 0.05 & \tabularnewline
 &  &  &  &  &  &  &  &  &  &  & \tabularnewline
\hline 
\end{tabular}
\caption{The average values of the center-of-mass phonon occupation over a time duration of 40 ms for the three-spin evolution revealed using an all-spin-parallel initial state and the two-spin evolution revealed using a mixed-spin initial state. The evolution occurs under the single-mode two-drive Hamiltonian $H_{\rm 2drive}^{\rm com}$ in Eq.~(\ref{eq:H2pair}) for the detunings and Rabi frequencies used in Figs.~\ref{fig:com3body} and ~\ref{fig:com2body}. The Hilbert space is restricted by truncating the phonon occupation of the center-of-mass mode at 6.}
\label{tab:phononcomI}
\end{table}
\begin{table}[b!]
\centering
\begin{tabular}{cccccc}
\hline 
 &  &  &  &  & \tabularnewline
 & $\frac{1}{2\pi}\{\delta,\Omega_{r}\}$[kHz] &  &  & $\{4,16\}$ & \tabularnewline
 &  &  &  &  & \tabularnewline
\hline 
 &  &  &  &  & \tabularnewline
 & $\left|\downarrow\downarrow\downarrow\right\rangle \otimes\left|0\right\rangle $ &  &  & 0.05 & \tabularnewline
 &  &  &  &  & \tabularnewline
 & $\left|\downarrow\uparrow\downarrow\right\rangle \otimes\left|0\right\rangle $ &  &  & 0.05 & \tabularnewline
 &  &  &  &  & \tabularnewline
\hline 
\end{tabular}
\caption{The average values of the center-of-mass phonon occupation over a time duration of 40 ms for the three- and two-body evolutions revealed by different initial states under the single-mode single-drive Hamiltonian $H_{\rm 1drive}^{\rm com}$ in Eq.~(\ref{eq:H1pair}) for the parameters considered in Fig.~\ref{fig:singlevsdouble_drives}. The Hilbert space is restricted by truncating the phonon occupation of the center-of-mass mode at 6.}
\label{tab:phononcomII}
\end{table}

As a result of these considerations, one might conclude that a larger detuning is the better choice. However, the anticipation from the more realistic multi-mode scheme is that a small detuning of the center-of-mass mode, together with a large phonon bandwidth, will be a feasible strategy to suppress the undesired effect of the other modes. In particular, within the two-drive scheme these modes spoil the cancellation of the two-spin dynamics. The optimal set of parameters for an experimental implementation should, therefore, be concluded from the analysis of the multi-mode scenario, as will be presented next.

\subsubsection{Multi-mode simulation}
\begin{figure*}[t!]
\centering
\includegraphics[width=2.05\columnwidth]{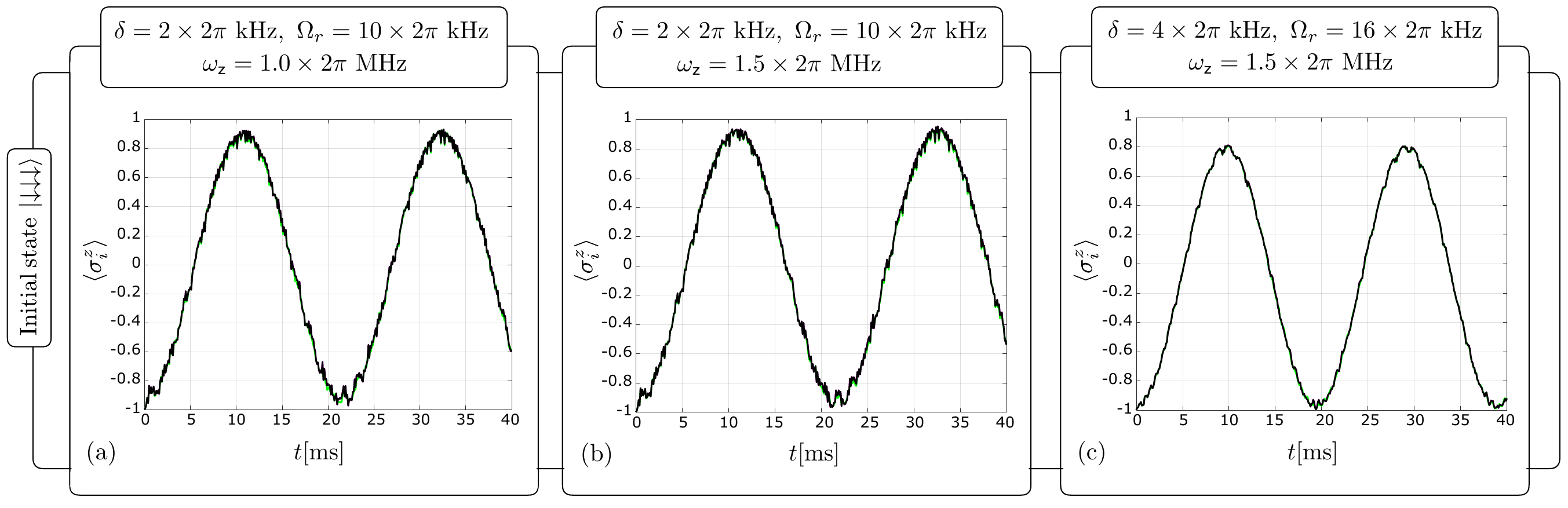}
\caption{Evolution of $\expval{\sigma_i^z}$ under the multi-mode Hamiltonian $H_{\rm 2drive}^{\rm multi}$ in Eq.~(\ref{eq:multimode}) from an initial state $\ket{\downarrow \downarrow \downarrow}\otimes\ket{0,0,0}$. The evolution of the spins  probes the system's three-spin interactions. Different panels consider different choices of detuning $\delta$, Rabi frequency $\Omega_r$, and axial trap frequency $\omega_{\mathsf{z}}$. Different colors (black, green, purple) correspond to the evolution of each of the different spins $(i=1,2,3)$, which, in contrast to the single-mode case, take different numerical values, but still can barely be discerned from one another.}
\label{fig:multi1}
\end{figure*}
\begin{figure*}[t!]
\centering
\includegraphics[width=2.05\columnwidth]{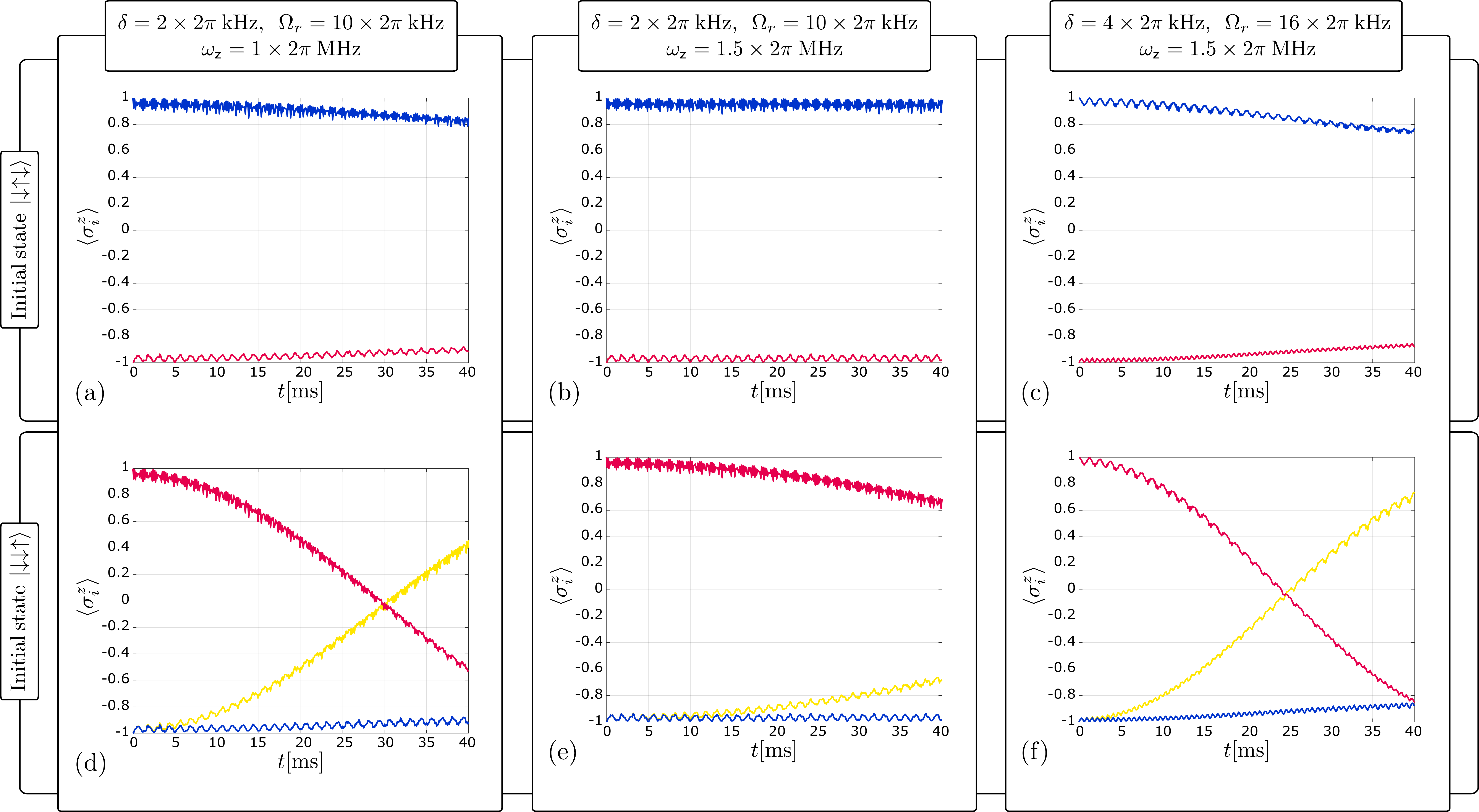}
\caption{Evolution of $\expval{\sigma_i^z}$ under the multi-mode Hamiltonian $H_{\rm 2drive}^{\rm multi}$ in Eq.~(\ref{eq:multimode}) from the initial state $\ket{\downarrow \uparrow \downarrow}\otimes\ket{0,0,0}$. (first row) and from the initial state $\ket{\downarrow \downarrow \uparrow}\otimes\ket{0,0,0}$ (second row). The evolution of the spins from these initial states probes the system's two-spin interactions. Different columns consider different choices of detuning $\delta$, Rabi frequency $\Omega_r$, and axial trap frequency $\omega_{\mathsf{z}}$. In each panel, different colors (yellow, blue, red) correspond to the evolution of each of the different spins $(i=1,2,3)$. In the case of a parity-symmetric initial state (first row), the yellow curves ($i=1$) is hidden by the red curve ($i=3$).}
\label{fig:multi}
\end{figure*}
Next, let us turn our attention to the multi-mode Hamiltonian, which up to second order in the Lamb-Dicke parameter reads
\begin{align}
H_{\rm 2drive}^{\rm multi}=&\sum_{i=1}^N\sum_{\mathsf{m}=1}^N\bigg[\frac{i}{2}\eta_{\mathsf{m},i} e^{i\omega_{\mathsf{m}}t}
\big(\Omega_r e^{i\mu_r t}+ 
\Omega_r' e^{i\mu_r' t} \big)
a^\dagger_{\mathsf{m}}\sigma^+_i
\nonumber \\
&\hspace{1.5 cm} -\frac{1}{4}\sum_{\mathsf{n}=1}^N \eta_{\mathsf{m},i} \eta_{\mathsf{n},i}\big( \Omega_b e^{i\mu_b t} + \Omega_b' e^{i \mu_b' t} \big)
\nonumber \\ 
& \hspace{1.5 cm} \times e^{-i(\omega_{\mathsf{m}}+\omega_{\mathsf{n}})t} \, a_{\mathsf{m}}a_{\mathsf{n}}\sigma^+_i \bigg]+{\rm h.c.},
\label{eq:multimode}
\end{align}
Here, $\Omega_r'$, $\Omega_b'$, $\mu_r'$, and $\mu_b'$ concern the second Raman-beam pair. They are defined in Eq.~(\ref{eq:primeparam}) in terms of the parameter $q$.

Simulations of the evolution under the multi-mode Hamiltonian were performed under the same conditions as for the single-mode case, i.e., with $q=1.3$, $\omega_{\rm rec}=26 \times 2\pi$~kHz, and $N=3$ ions. In addition to the transverse trap frequency, $\omega_{\mathsf{x}}=5 \times 2\pi$ MHz, the axial trap frequency $\omega_{\mathsf{z}}$ also needs to be fixed, as it determines the bandwidth of the phonon spectrum, i.e., transverse modes separations. The results for different $\omega_{\mathsf{z}}$ values are shown in Fig.~\ref{fig:multi1} for an all-spin-parallel initial spin state, and in Fig.~\ref{fig:multi} for different mixed-spin initial states. For the detuning and the Rabi frequencies, those parameters which worked well in the single-mode approximation have been chosen, i.e., $\delta=2 \times 2\pi$~kHz with $\Omega_r=10 \times 2\pi$~kHz and $\delta=4 \times 2\pi$~kHz with $\Omega_r=16 \times 2\pi$~kHz.

From an initial state $\ket{\downarrow \downarrow \downarrow} \otimes \ket{0,0,0}$ (where the latter ket corresponds to the vacuum of three phonon modes), one observes three-spin oscillations with good fidelity, as shown in Fig.~\ref{fig:multi1}(a)-(c). This is not surprising, since two-spin interactions are automatically suppressed by the choice of initial state, and off-resonant contributions from the other modes are negligible due to their large detuning. Notably, the evolution of the three ions is almost exactly the same in these cases, that is, one can hardly discern the black curve in all panels of Fig.~\ref{fig:multi1} for ion 1, from the green and the purple curves for ion 2 and ion 3. 

An interesting observation is that the oscillation amplitude when evolving under the multi-mode scenario is larger than that for the single-mode approximation with similar parameters (compare e.g., Fig. \ref{fig:com3body}(b) and Fig. \ref{fig:multi1}(b)).
This might seem counterintuitive given the exact cancellation of single- and two-spin dynamics in the single-mode approximation. An explanation of this behavior could be a destructive interference of the two factors which contribute to the suppression of the oscillations' amplitude hence decrease in fidelity: i) the effective single-spin terms in Eq.~\ref{eq:H1sigmamulti}, and ii) noise, i.e., any terms which are not accounted for by the effective Hamiltonian, that involve various spin-phonon couplings. In the single-mode calculation, only the latter contribution appears as the former is fully suppressed in the two-drive scheme. In the multi-mode case, both factors contribute, and it is plausible that they can partially cancel each other out.

Let us now probe the two-body dynamics by initializing the system in a mixed-spin state. Since the second pair of Raman beams is designed to only cancel the two-spin terms stemming from the center-of-mass mode, one observes an enhanced amount of two-spin dynamics in the multi-mode scenario, but one also has to differentiate between parity-symmetric initial states, like $|\!\!\downarrow \uparrow \downarrow \rangle$ shown in Fig.~\ref{fig:multi}{(a)-(c)}, and asymmetric initial states like $|\!\!\uparrow \downarrow \downarrow\rangle$, shown in Fig.~\ref{fig:multi}{(d)-(f)}. The ``tilt'' mode induces two-spin interactions only if the first and the third spin are different, and therefore this mode, which is next to the center-of-mass mode in the phonon spectrum, does not contribute in the parity-symmetric case. Therefore, the two-spin dynamics in Fig.~\ref{fig:multi}{(a)-(c)} remains strongly suppressed. However, this is not the case for asymmetric initial states, and much faster two-spin dynamics is observed, see Fig.~\ref{fig:multi}{(d)-(f)}.

Nevertheless, by increasing the bandwidth of the phonon spectrum, the detuning of the modes other than the center-of-mass mode can be increased, and thereby their effect will decrease. This strategy works best if the detuning of the center-of-mass mode is kept small. To enhance the phonon bandwidth, the axial confinement can be increased, e.g. by changing $\omega_{\mathsf{z}}$ from $1 \times 2\pi$ MHz to $1.5 \times 2\pi$ MHz, see Figs.~\ref{fig:multi1} and \ref{fig:multi}. While this change has almost no effect on the three-spin evolution, the two-spin dynamics is significantly slowed down. 

Moreover, from comparison between panels (b)-(e) and (c)-(f) of Fig.~\ref{fig:multi}, one can recognize the effect of an increase of the detuning from 2~kHz to 4~kHz (together with a proportional increase of the Rabi frequency to keep the three-spin dynamics similarly fast in both cases). This increase of detuning, while slightly improving the fidelity of the three-spin evolution (as already seen in the single-mode case), leads to significantly faster two-spin dynamics. As anticipated earlier, this shows that in order to achieve a quantum simulator with purely three-spin interactions, one must keep the detuning as small as possible.

For completeness, the average phonon number over a duration of 40 ms for the different combinations of parameters and initial states as used in Figs.~\ref{fig:multi1} and \ref{fig:multi} are reported in Table~\ref{tab:phononcomIII}. As is seen, the average phonon occupation remains well below one. Furthermore, not surprisingly in all cases the most occupied mode is the center-of-mass mode, as it the beams are more closely tuned to the center-of-mass mode than the rest.
\begin{table}[t!]
\centering
\begin{tabular}{cccccccc}
\hline 
 &  &  &  &  &  &  & \tabularnewline
$\frac{1}{2\pi}\{\delta,\Omega_{r},\omega_{\mathsf{z}}\}$[kHz] &  &  & $\{2,10,1000\}$ &  & $\{2,10,1500\}$ &  & $\{4,16,1500\}$\tabularnewline
 &  &  &  &  &  &  & \tabularnewline
\hline 
 &  &  &  &  &  &  & \tabularnewline
$\left|\downarrow\downarrow\downarrow\right\rangle \otimes\left|0,0,0\right\rangle $ &  &  & 0.20 &  & 0.19 &  & 0.08\tabularnewline
 &  &  &  &  &  &  & \tabularnewline
$\left|\downarrow\uparrow\downarrow\right\rangle \otimes\left|0,0,0\right\rangle $ &  &  & 0.07 &  & 0.08 &  & 0.04\tabularnewline
 &  &  &  &  &  &  & \tabularnewline
$\left|\downarrow\downarrow\uparrow\right\rangle \otimes\left|0,0,0\right\rangle $ &  &  & 0.09 &  & 0.08 &  & 0.05\tabularnewline
 &  &  &  &  &  &  & \tabularnewline
\hline 
\end{tabular}
\caption{The average values of the total phonon occupation in all modes over a time duration of 40 ms for the three and two-body evolutions under the multi-mode Hamiltonian $H_{\rm 2drive}^{\rm multi}$ in Eq.~(\ref{eq:multimode}). In all cases, the average values for occupation of the non-center-of-mass modes are less than $0.01$. The Hilbert space is restricted by truncating the phonon occupation of the center-of-mass mode at 6. The phonon occupation of the other two modes is truncated at 2.}
\label{tab:phononcomIII}
\end{table}

In summary, the implementation of the three-spin Hamiltonian is possible with an appropriate choice of the experimental parameters. Mainly, this requires increasing the axial trap frequency as much as possible, as well as keeping the detuning small. As an optimal lower limit for the detuning, the simulation carried out in this section determines the value $\delta\approx 2 \times 2\pi$~kHz. Limitations to the axial frequency are more general, given by zig-zag deformation of the ion chain, which happens when $\omega_{\mathsf{z}} > \frac{\omega_{\mathsf{x}}}{0.73} N^{-0.86}$~\cite{Steane1997}. For the case of $N=3$ and the transverse frequency of this simulation, the bound on the allowed axial frequency is $\approx 2.67$ MHz.

\subsubsection{Preparation of a GHZ state}
To conclude this section, we consider the possibility of using the three-spin coupling scheme to prepare a maximally entangled state of three spins. Specifically, by subjecting an all-parallel spin state to the three-spin Hamiltonian for the proper amount of time, i.e., $\approx \pi/(2J_3^{\rm 2drive})$, a Greenberger–Horne–Zeilinger (GHZ)-like state $|{\rm GHZ}(\varphi)\rangle=\frac{1}{\sqrt{2}}(|\uparrow \uparrow \uparrow \rangle + e^{i\varphi} |\downarrow \downarrow \downarrow \rangle)$ can be obtained, where in our scheme the phase angle is $\varphi=\frac{3\pi}{2}$.\footnote{To set the relative phase of the two states equal to unity, a trivial single-qubit phase gate can be applied to one of the ions at the beginning of the operation.}

\begin{figure}[t!]
\centering
\includegraphics[width=0.75\columnwidth]{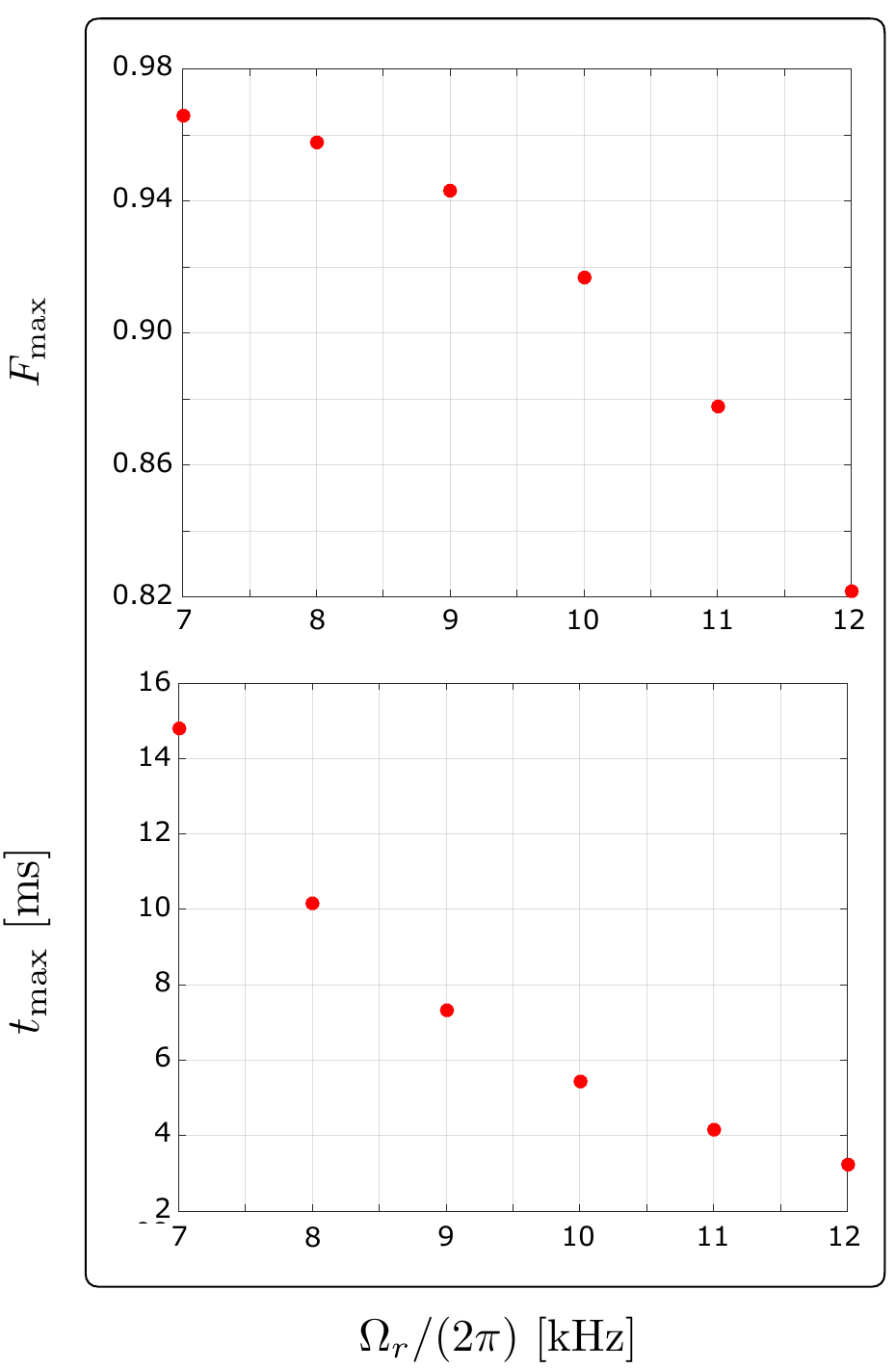}
\caption{Maximum fidelity for the GHZ-state preparation, along with the time to reach such a fidelity, for different values of Rabi frequency $\Omega_r$, at a fixed detuning $\delta=2\times 2\pi~\rm{kHz}$ and axial frequency $\omega_{z}=1500 \times 2\pi~\rm{kHz}$. The GHZ state considered here is defined as $\ket{\rm{GHZ}(\frac{3\pi}{2})}=\frac{1}{\sqrt{2}}\left(\ket{\downarrow\downarrow\downarrow}-i\ket{\uparrow\uparrow\uparrow}\right)$, and the phonon degrees of freedom are traced out in the intermediate-states density matrices.}
\label{fig:fidelity}
\end{figure}

To determine how well this GHZ state is reached, one can calculate the fidelity $F$ as the system evolves defined as
\begin{eqnarray}
F(t) \equiv \left[{\rm tr} \sqrt{\sqrt{\rho_{\rm GHZ}} \, \rho(t) \, \sqrt{\rho_{\rm GHZ}}}\right]^2,
\end{eqnarray}
where $\rho_{\rm GHZ}$ denotes the density matrix corresponding to the state $|{\rm GHZ}(\frac{3\pi}{2})\rangle$. Since this is a pure state, the simplified relation $F(t)=\braket{{\rm GHZ}(\frac{3\pi}{2})|\rho(t)|{\rm GHZ}(\frac{3\pi}{2})}$ can also be used. $\rho(t)$ denotes the reduced density matrix of the time-evolved state after tracing out the phonon degrees of freedom. The maximum fidelity $F_{\rm max}$ is plotted in Fig.~\ref{fig:fidelity} for different choices of the Rabi frequency $\Omega_{\rm r}$. We also plot the time $t_{\rm max}$ at which the maximum fidelity is reached. It can be seen that a fidelity $>0.96$ can be reached within a time less than 10~ms for $\Omega_{\rm r}=8 \times 2\pi$ kHz. Slight increases of the Rabi frequencies allow to reduce the preparation time to smaller than 5 ms, keeping $F_{\rm max}$ still as large as approximately 0.92. A comparison of the fidelities achieved in digital universal implementations of quantum circuits using only two-qubit entangling gates compared with the three-qubit entangling operation proposed here will be discussed more closely in Sec.~\ref{sec:digital}.

\section{Applications in quantum simulation: a lattice gauge theory example
\label{sec:schwinger}}
\noindent
Quantum simulation of quantum many-body systems can potentially benefit from a direct implementation of three-spin interactions. Examples include simulating certain lattice gauge theories whose dynamics can be mapped to multi-spin Hamiltonians~\cite{chandrasekharan1997quantum, sala2018variational}, as well as spin systems with two- and three-spin interactions which can exhibit nontrivial phase diagrams~\cite{pollmann2010entanglement, verresen2017one, verresen2018topology, smith2019crossing, tseng1999quantum, wannier1950antiferromagnetism, peng2009quantum, d1998level, penson1988conformal, gross1984simplest, GARDNER1985747, nieuwenhuizen1998quantum, tsomokos2008chiral, Wen2004}. Among the features of the scheme of this work that can be useful in simulating such physical systems are the possibility of a direct simulation of the three-spin dynamics without the need for digitalization of the evolution, as well as the ability to tune the relative strength of two- and three-spin interactions. In this section, an example in the context of lattice gauge theory will be presented.\footnote{We refer the reader to Refs.~\cite{pollmann2010entanglement, verresen2017one, verresen2018topology, smith2019crossing, tseng1999quantum, wannier1950antiferromagnetism, peng2009quantum, d1998level, penson1988conformal, gross1984simplest, GARDNER1985747, nieuwenhuizen1998quantum, tsomokos2008chiral, Wen2004} for applications in the context of condensed-matter theory.} This section also investigates the benefit of implementing three-qubit gates in place of two-qubit gates to achieve higher-fidelity decomposition of certain time-evolution operators in quantum simulation.

\subsection{A U(1) quantum link model with  the three-spin coupling scheme
\label{sec:QLM}}
\begin{figure*}[t!]
\centering
\includegraphics[scale=0.65]{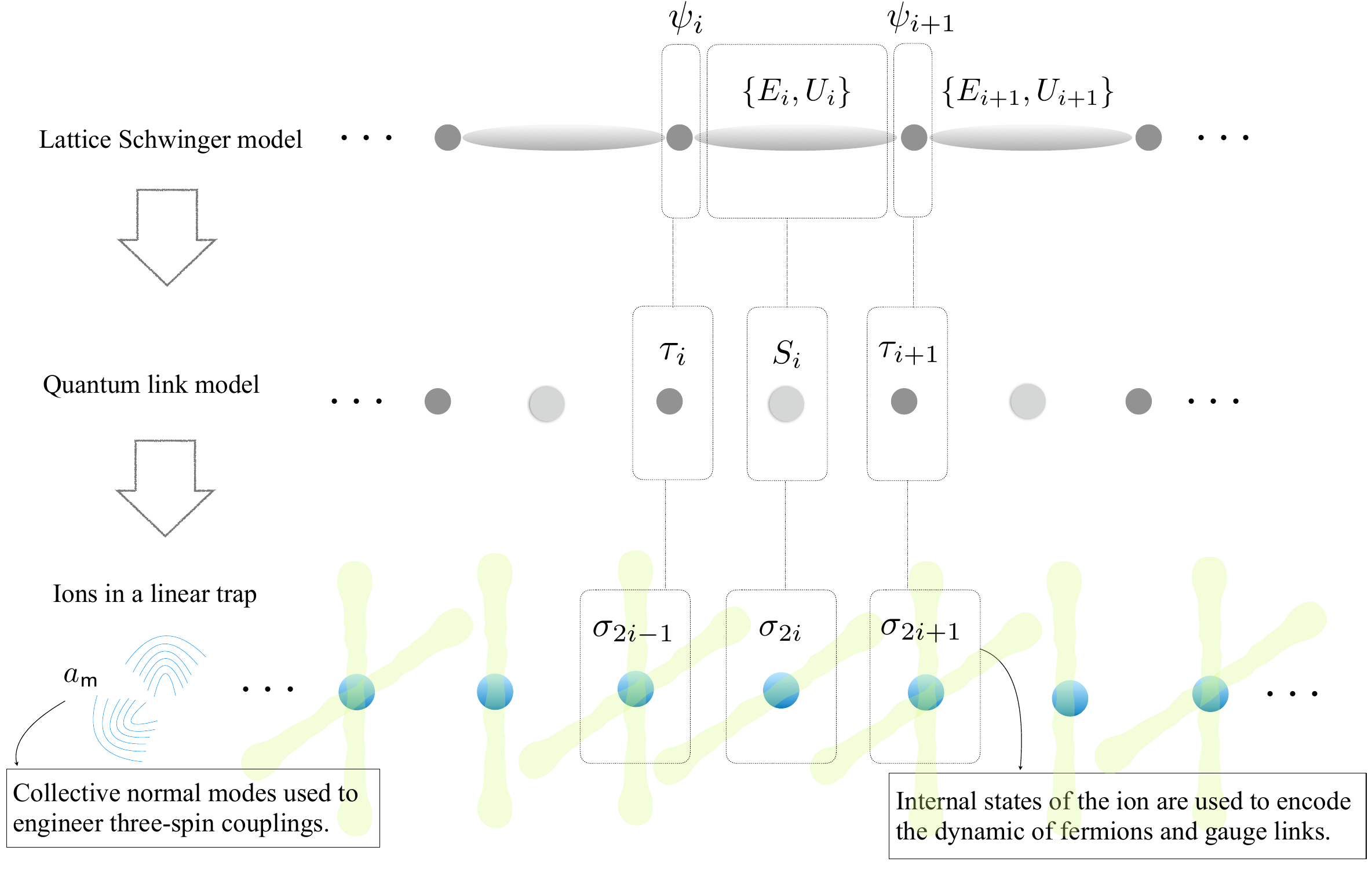}
\caption{
The mapping of the degrees of freedom of the lattice Schwinger model to a quantum link model, that is ultimately mapped to (quasi)spin degrees of freedom of a linear chain of trapped ions, which are addressed with given sets of laser beams as described in the text. The ion-laser interactions induce an effective three-spin Hamiltonian via virtually-excited phonons.
}
\label{fig:qlm-mapping}
\end{figure*}
The U(1) lattice gauge theory in 1+1 dimensions, i.e., the lattice Schwinger model, is a valuable prototype of QCD and has long served as a testbed for benchmarking quantum simulation of lattice gauge theories on quantum hardware~\cite{Martinez:2016yna, Klco:2018kyo, Mil:2019pbt, Yang:2020yer}. In the Kogut-Susskind formulation~\cite{kogut1975hamiltonian}, the model consists of fermion-gauge interactions, electric-field contribution, and a staggered fermion mass term:
\begin{eqnarray}
&&H_{\rm U(1)} = x \sum_{i=1}^{N_{\rm stag}-1}\big[ \psi_i^\dagger U_i \psi_{i+1} + {\rm h.c.} \big]+
\sum_{i=1}^{N_{\rm stag}-1} E_i^2+
\nonumber\\
&& \hspace{4.8 cm}\mu \sum_{i=1}^{N_{\rm stag}} (-1)^i \psi_i^\dagger \psi_i.
\end{eqnarray}
Fermions $\psi_i$ sit on site $i$, while the conjugate-variable pairs $\{E_i,U_i\}$ belong to the link originating from site $i$, see Fig.~\ref{fig:qlm-mapping}. $N_{\rm stag}$ denotes the number of staggered sites. Rescaled dimensionless couplings $x$ and $\mu$\footnote{Not to be confused with the lasers' frequency in the previous section.} are used such that the Hamiltonian is expressed in dimensionless unit. Open boundary conditions will be considered such that the electric-field flux into the lattice, $E_0$, is fixed. The continuum limit of the expectation values of the observable $O$ for a fixed mass over coupling ratio is obtained via a double-ordered limit: $\lim_{x \to \infty} \lim_{N_{\rm stag} \to \infty} \braket{O}$, a limit that will not be considered here. The states are characterized by bosonic and fermionic quantum numbers. Explicitly, bosonic quantum number $n^{(g)}_i$ is associated with the discrete spectrum of a quantum rotor satisfying $E_i\ket{n^{(g)}_i}=n^{(g)}_i\ket{n^{(g)}_i}$ and $U_i\ket{n^{(g)}_i}=\ket{n^{(g)}_i+1}$ with $n^{(g)}_i\in \mathbb{Z}$. The fermionic quantum number $n^{(f)}_i$ is associated with the fermionic occupation at each site, which satisfies $n^{(f)}_i \in \{0,1\}$. An unoccupied (occupied) odd (even) site represents the presence of an electron (positron). The physical states are those annihilated by the Gauss's law operator $G_j=E_j-E_{j-1}-\psi_j^\dagger \psi_j +\frac{1}{2}\big[1-(-1)^j \big]$ at each site. 

A finite-dimensional formulation of the same U(1) theory is described by a quantum link model~\cite{chandrasekharan1997quantum, Brower:1997ha}. Within this formulation, and upon performing a Jordan-Wigner transformation of the staggered fermions, the Hamiltonian becomes 
\begin{eqnarray}
&&H_{\rm QLM} = J \sum_{i=1}^{N_{\rm stag}-1}\big[ \tau_i^+ S_i^+ \tau_{i+1}^- + {\rm h.c.} \big]+\sum_{i=1}^{N_{\rm stag}-1}S_z^2+
\nonumber\\
&& \hspace{5.1 cm} \mu \sum_{i=1}^{N_{\rm stag}} (-1)^i \sigma_i^z,
\label{eq:Hqlm}
\end{eqnarray}
where $\tau_i$ are Pauli matrices at each site, and $S$ is a spin operator. While the exact Schwinger model is recovered in the limit of $S \to \infty$, for low-energy quantities, small $S$ values provide accurate approximations to the exact theory, as shown in e.g., Ref.~\cite{zache2021achieving}.\footnote{Note that only in the limit of $S\to\infty$, $J$ approaches the original coupling $x$, see Ref.~\cite{zache2021achieving}.} To map this Hamiltonian to the trapped-ion effective Hamiltonian, the spin-$\frac{1}{2}$ limit of the quantum links will be considered, and $S_i$ in Eq.~(\ref{eq:Hqlm}) will be replaced by the Pauli operators at each site. The electric-field contribution is then trivial and one is left with matter-gauge interaction in form of a nearest neighbor three-spin operator, as well as the staggered mass term.

Such a Hamiltonian can be mapped to the effective Hamiltonian of the trapped-ion simulator. Explicitly,
\begin{equation}
H_{\rm Ion} = J \sum_{i=1}^{N_{\rm stag}-1
}\big[ \sigma_{2i-1}^+ \sigma_{2i}^+ \sigma_{2i+1}^- + {\rm h.c.} \big]+
\mu \sum_{i=1}^{N_{\rm stag}} (-1)^i \sigma_{2i-1}^z.
\label{eq:HqlmIon}
\end{equation}
Note that $N=2N_{\rm stag}-1$ ions are needed to encode the dynamics with open boundary conditions. The basis states are simply the direct product of eigenstates of the Pauli operator $\sigma^z$, while the physical states are those annihilated by the Gauss's law operator, which in this formulation reads: 
$G_i=\frac{1}{2} \left[ \sigma^z_{2i}-\sigma^z_{2i-2}-\sigma_{2i-1}^z-(-1)^{i}\right]$ with $\sigma^z_{0}$ fixed by (open) boundary conditions. A diagrammatic representation of the mapping between degrees of freedom in the original lattice Schwinger model, the quantum link model, and the trapped-ion simulator is shown in Fig.~\ref{fig:qlm-mapping}.

In order to apply the scheme of the previous section to simulate the dynamics governed by Eq.~(\ref{eq:HqlmIon}), one first needs to perform a local unitary transformation consisting of $\pi$-rotations on every other odd-even pairs of ions, as well as on the last ion in the chain. Explicitly,
\begin{equation}
\mathcal{U} \equiv \sigma^x_{2N_{\rm stag}-1}\prod_{i=1}^{\frac{N_{\rm stag}}{2}-1}\sigma^x_{4i-1}\sigma^x_{4i}
\end{equation}
is applied at the beginning and the end of the simulation. The Hamiltonian in the transformed basis is
\begin{equation}
H^{(\rm rot.)}_{\rm Ion} = J \sum_{i=1}^{N_{\rm stag}-1
}\big[ \sigma_{2i-1}^+ \sigma_{2i}^+ \sigma_{2i+1}^+ + {\rm h.c.} \big]-
\mu \sum_{i=1}^{N_{\rm stag}} \sigma_{2i-1}^z.
\label{eq:HqlmIonRot}
\end{equation}
For an analog simulation of the dynamics governed by this Hamiltonian, one can use the two-drive scheme discussed in the previous section, in addition to a Stark-shift beam to induce the evolution under the mass term. Specifically, the gauge-matter term proportional to three-spin interactions can be mapped to the effective Hamiltonian in Eq.~(\ref{eq:H3sigmamulti}). Acting on three ions, the two-drive scheme produces the desired three-spin interactions, while suppressing single- and two-spin terms. For $N \geq 3$, however, the ions at locations $2i+1$ for $i \in \{1,\cdots,N_{\rm stag}-2\}$ need to be addressed with two sets of Raman beams, as these ions participate in three-spin interactions with their left and right neighbors simultaneously. This can be achieved through local addressing of the ions, hence individual control of Rabi frequencies. Besides the need for individual addressing, one needs to mitigate the problem of deviating from the effective three-spin dynamics given multiple drives on each third ion in the chain. One strategy is to apply lasers that address different transverse modes of the chain in an alternate pattern. Explicitly, ions $\{2i-1,2i,2i+1\}$ can be addressed by beams with $\Delta \bm{k}=\Delta k \hat{\bm{\mathsf{x}}}$ while ions $\{2i+1,2i+2,2i+3\}$ can be addressed with beams with $\Delta \bm{k}=\Delta k \hat{\bm{\mathsf{y}}}$, for $i\in \{1,3,\cdots,N_{\rm stag}-3\}$, see Fig.~\ref{fig:qlm-mapping}.\footnote{With open boundary conditions, ions $1$ and $2N_{\rm stag}-1$ do not need to be addressed with two sets of orthogonal Raman pairs.}

In this scheme, there will still be two types of undesired contributions to the dynamics even with two sets of alternate drives. First, there will be terms proportional to phonon creation and annihilation operators in both modes, that go as $e^{i(\delta-\tilde{\delta})t}$ or slower, where $\delta$ and $\tilde{\delta}$ are the detuning from the center-of-mass frequency of each set of the modes. Thus, multiple drives with similar detunings on a single ion will give rise to undesired spin-phonon entanglement. These contributions can be made off-resonant by setting the two detunings sufficiently differently, and by compensating for the required equal couplings by adjusting the corresponding Rabi frequencies. The second type of terms can result from cross-coupling among e.g., ions in the sets $\{2i-1,2i,2i+1\}$ and $\{2i+3,2i+4,2i+5\}$, as the beams aligned with the same principal axis of the trap are used for the two non-adjacent sets. Such cross-couplings can result in long-range effective spin interactions along the chain. They can be made off-resonant by adopting different detunings for each set of the beams. To increase the distance among the ions addressed by lasers along the same principal axes, axial modes of motion can be addressed too, but these modes are lower in frequency, and the second sideband drive with respect to the center-of-mass mode would unavoidably induce the first sideband transitions of several modes other than the center-of-mass mode, which complicates the effective dynamics. Increasing the axial trapping frequency, as well as detuning close the highest frequency (``zig-zag'') axial mode, can help with this limitation.

Another complication with scaling up the scheme of the previous section to simulate a large instance of the quantum link model is related to the proliferation of the normal modes and the increased contribution from modes other that the center-of-mass mode to the dynamics---an effect that is anticipated to worsen the cancellation of the undesired single- and two-spin contributions. As a result, the fidelity of the simulation compared with the exact effective three-spin dynamics degrades as a function of the number of the trapped ions. As performing a numerical study of the full dynamics to estimate the expected outcome is costly and reaches the limits of classical methods quickly, experimental implementations on the quantum simulator are the only means by which to assess the performance of the scheme for larger system sizes.

The mass term can be implemented by applying, simultaneously with the lasers that induce the three-spin coupling, a Stark shift on participating ions, i.e., ions $2i-1$ for $i \in \{1,\cdots,N_{\rm stag}\}$, corresponding to a longitudinal effective magnetic field $B_z$ applied to the ions. Alternatively, properly shifted red and blue sideband transitions can effectively implement such a magnetic-field interaction, as described, e.g., in Refs.~\cite{wall2017boson, Davoudi:2019bhy, monroe2021programmable}. However, arbitrarily large masses cannot be simulated if long evolution times are desired, as undesired cross-coupling terms with the lasers implementing the three-spin coupling occurs. By requiring the model parameters to match those in experiment for the effective magnetic-field term throughout the evolution, that is by setting $\mu t_{\rm QLM} = B_z t$, the undesired terms will be small for $|B_z t \eta \Omega_r/\delta| = |\mu t_{\rm QLM} \eta \Omega_r/\delta| \ll 1$ and $|B_z t \eta^2 \Omega_b/\delta| = |\mu t_{\rm QLM} \eta^2 \Omega_b/\delta| \ll 1$.
\begin{figure*}[t!]
\centering
\includegraphics[scale=0.95]{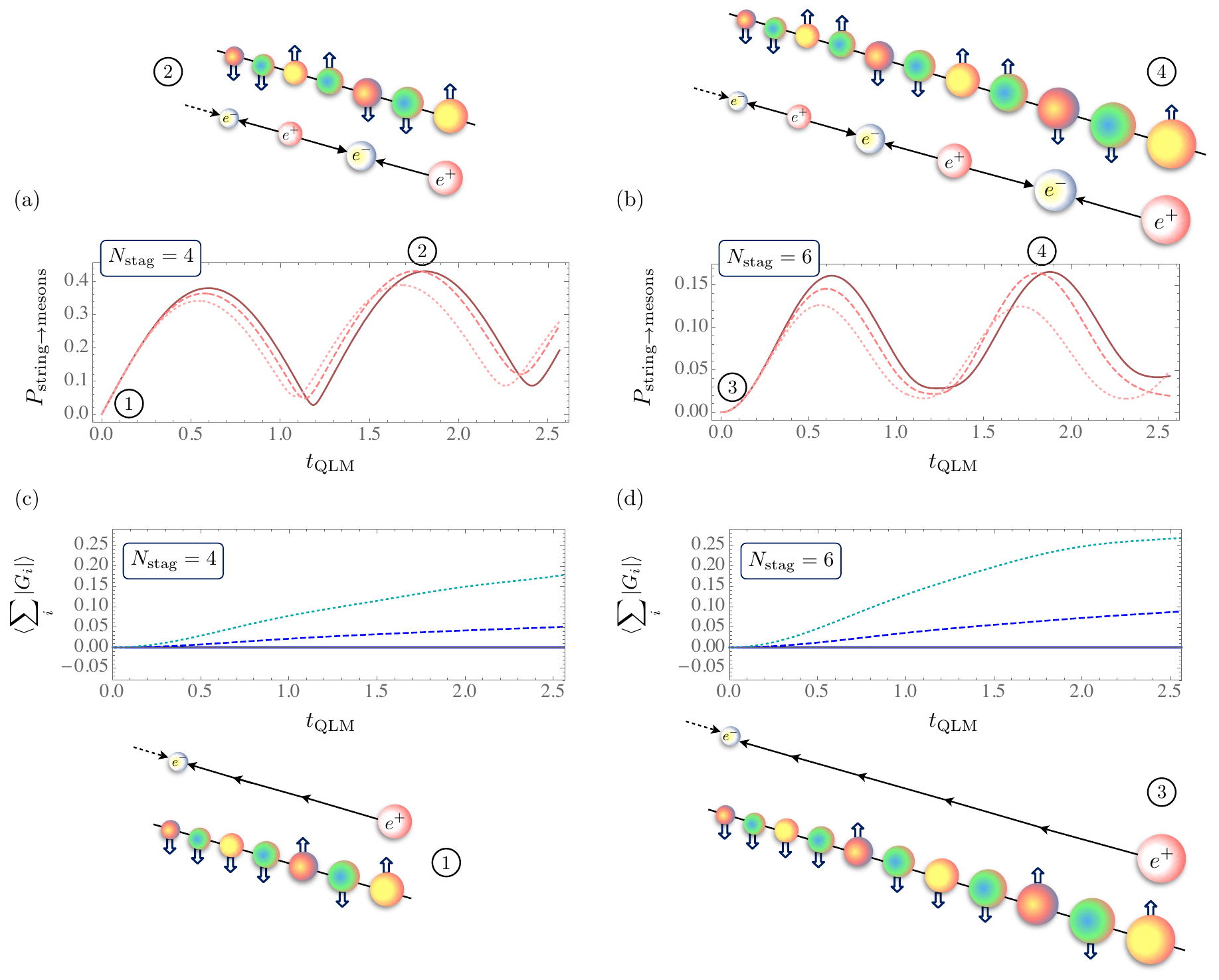}
\caption{(a) The (absolute value) of the overlap between a time-evolved string state $\ket{\psi_{\rm str}}$ and a fully-occupied mesonic state $\ket{\psi_{\rm mes}}$, $P_{{\rm string} \to {\rm mesons} } \equiv |\braket{\psi_{\rm mes}|e^{-iH_{\rm QLM}t_{\rm QLM}}|\psi_{\rm str}}|$, as a function of the (scaled dimensionless) time $t_{\rm QLM}$ for a lattice with $N_{\rm stag}=4$ fermion sites, corresponding to $N=7$ ion sites. For the dashed and dotted curves, the Hamiltonian must be replaced with the non-exact $H_{\rm QLM}^\prime$ and $H_{\rm QLM}^{\prime\prime}$ Hamiltonians, respectively. (b) The same quantity plotted for $N_{\rm stag}=6$ fermion sites, corresponding to $N=11$ ion sites. The graphical representations of states, both in terms of the electron, positron, and electric-field strings and in terms of the (quasi)spins of each corresponding ions are shown for the initial string state, and for a fully-occupied mesonic state whose probability amplitude is maximum at the points denoted. (c) The expectation value of the lattice sum of the (absolute value of) the Gauss's law operator between a time-evolved string state, $\braket{\sum_i|G_i|} \equiv \braket{\psi_{\rm str}|e^{iH_{\rm QLM}t_{\rm QLM}}\frac{1}{2N_{\rm stag}-3}\sum_{i=1}^{N_{\rm stag}-1}|G_i| e^{-iH_{\rm QLM}t_{\rm QLM}}|\psi_{\rm str}}$ for $N_{\rm stag}=4$ fermion sites, corresponding to $N=7$ ion sites. For the dashed and dotted curves, the Hamiltonian must be replaced with the non-exact $H_{\rm QLM}^\prime$ and $H_{\rm QLM}^{\prime\prime}$ Hamiltonians, respectively. (d) The same quantity as in (c) for $N_{\rm stag}=6$ fermion sites, corresponding to $N=11$ ion sites. The maximum breakdown of the Gauss's law corresponds to $\braket{\sum_i|G_i|}=1$.
}
\label{fig:string-breaking}
\end{figure*}
%

\subsection{A numerical study of inexact dynamics with a crude model
\label{sec:QLM-numerics}}
It is interesting to investigate if presently-available trapped-ion simulators can reveal nontrivial nonperturbative features of strong dynamics in the lattice Schwinger model despite the imperfect fidelity of implementing effective three-spin dynamics as quantified in our numerical study. Let us consider a 7-ion and an 11-ion system, as these system sizes are currently available in both digital and analog modes (see e.g., Refs.~\cite{wright2019benchmarking, egan2020fault, bruzewicz2019trapped, monroe2021programmable}). These systems can encode a quantum link model with $N_{\rm stag}=4$ and $N_{\rm stag}=6$ staggered matter sites, respectively. The numerical study of Sec.~\ref{sec:numerics} with realistic experimental parameters established that the period of the undesired two-spin evolution is at worse $\sim 20$ times larger than that of the desired three-spin evolution, considering contributions from all the modes. This indicates an effective three-spin coupling that is at least $\sim 20$ time larger than the two- (and one-) spin couplings---a consequence of the effectiveness of the two-drive scheme in suppressing the undesired contributions. The accuracy of this scheme, of course, drops as a function of time, particularly as a result of noise, coupling to  environment, and decoherence---effects that will be left out of our analysis and can only be carefully quantified in an experimental implementation. An accurate prediction of the dynamics in the quantum link model with the full Hamiltonian requires inputting a comprehensive model of interactions including phonons in chains longer than three ions, and hence is beyond a simple numerical investigation. Nonetheless, one can still come up with a crude model of interactions to describe the quantum link model under an imperfect Hamiltonian that qualitatively resembles that of the full Hamiltonian. For this purpose, we consider the Hamiltonian in Eq.~(\ref{eq:Hqlm}) with the addition of an all-to-all two-spin interactions that are uniform in strength, and to be conservative, are taken to be 10 times and 5 times weaker than the desired three-spin interactions. Additionally, single-spin interactions on all ions are included to modify the mass term with uniform coefficients that are 10 and 5 times weaker than the true mass. Explicitly,
\begin{eqnarray}
&&H_{\rm QLM}^{\prime(\prime\prime)}=H_{\rm QLM}+\frac{x}{g^{\prime(\prime\prime)}} \sum_{\underset{j\neq k}{j,k=1}}^{2N_{\rm stag}-1
}\big[ \sigma_j^+ \sigma_k^- + {\rm h.c.} \big]+ 
\nonumber\\
&&\hspace{4.85 cm}\frac{\mu}{g^{\prime(\prime\prime)}} \sum_{j=1}^{2N_{\rm stag}-1} \sigma_j^z,
\label{eq:Hqlmp}
\label{eq:Hqlmpp}
\end{eqnarray}
with $g^\prime=10$ and $g^{\prime\prime}=5$. One can now ask, assuming this crude model captures the imperfection of the effective model obtained in Sec.~\ref{sec:derivation}, whether interesting nontrivial features of the quantum link model of the lattice Schwinger model, such as string breaking and pair creation after a quench, can be observed in experiment. Furthermore, given the gauge-symmetry violating interactions in play, will the degree of symmetry breaking in the evolved states be small with a gauge-invariant initial states, such that qualitative features of a constrained dynamics are not fully lost?

Figure~\ref{fig:string-breaking} shows the result  of one such study. The initial gauge-invariant state consisting of positive and negative electric charges at the far ends of lattice with an electric-field string (quantum-link) attaching them, the so-called ``string'' state, is evolved in time, and its overlap is measured with a fully-occupied ``mesonic'' state consisting of $N_{\rm stag}/2$ electron-positron pairs, connected by proper electric-field fluxes to satisfy Gauss's law. Explicitly, the quantity 
\begin{eqnarray}
P_{{\rm string} \to {\rm mesons}} \equiv |\braket{\psi_{\rm mes}|e^{-iH_{\rm QLM}t_{\rm QLM}}|\psi_{\rm str}}|
\end{eqnarray}
is calculated via exact diagonalization of the $N_{\rm stag}=4$ and $N_{\rm stag}=6$ site theories, where $\ket{\psi_{\rm str}}$ denotes the string state and $\ket{\psi_{\rm mes}}$ denotes the mesonic state. Besides the evolution with the exact Hamiltonian $H_{\rm QLM}$ (in solid curves), the two modified Hamiltonian $H_{\rm QLM}^\prime$ and $H_{\rm QLM}^{\prime\prime}$ are also considered (in dashed and dotted curves, respectively). The plots indicate that the generation of  a significant amplitude for a mesonic state out of an initial string state after a quantum quench is retained for imperfect Hamiltonians, and is only slowly diminished as function of time, particularly for the smaller perturbation. The reason for the decline in the meson-state amplitude is the leakage to the vast unphysical Hilbert space that can occur since the modified Hamiltonians do not respect the Gauss's law constraints.

To reveal the degree of gauge-symmetry violation in the nonperturbative dynamics, Fig.~\ref{fig:string-breaking} also depicts the quantity
\begin{eqnarray}
&&\braket{\sum_i|G_i|} \equiv \langle \psi_{\rm str}|e^{iH_{\rm QLM}t_{\rm QLM}}
\nonumber\\
&&\hspace{0.9  cm}\frac{1}{2N_{\rm stag}-3}\sum_{i=1}^{N_{\rm stag}-1}|G_i| e^{-iH_{\rm QLM}t_{\rm QLM}}|\psi_{\rm str} \rangle,
\end{eqnarray}
with the Gauss's law operator $G_i$ defined above in the spin-$\frac{1}{2}$ formulation of the quantum link model. For a gauge-invariant state, therefore, $\braket{\sum_i|G_i|}=0$, while for a state with maximum violation of the Gauss's law constraint, $\braket{\sum_i|G_i|}=1$.\footnote{Hence the normalization $\frac{1}{2N_{\rm stag}-3}$ adopted in the definition of $\braket{\sum_i|G_i|}$. Note that  with open boundary conditions, the last fermionic site is left open and is not counted toward the sum.} As is seen in the figure, for $H_{\rm QLM}^\prime$, the gauge-symmetry violation remains at a few-percent level at early times. For $H_{\rm QLM}^{\prime\prime}$, the violation can be significant, for both lattice sizes considered, although it remains well below one at early times. Despite the breakdown of gauge invariance, the qualitative behavior of the evolution under exact Hamiltonian appears to be robust with respect to the perturbations introduced (for a few meson-state revival cycles), making  this model a suitable first case study in the upcoming implementations.\footnote{The robustness of gauge-invariant dynamics in latice gauge theories is thoroughly studied in Refs.~\cite{Halimeh:2020djb, VanDamme:2021teo}.}

\subsection{Digital versus analog
\label{sec:digital}}
As demonstrated in Sec.~\ref{sec:numerics}, the three-spin(qubit) coupling scheme can be used to prepare a maximally-entangled three-qubit state, such as the GHZ state, with high fidelity. The two-qubit entangling gates, i.e., the conventional M{\o}lmer-S{\o}rensen (MS) gate, have achieved increasingly high fidelities ($\sim 0.995$ for a three-qubit system for comparison~\cite{bruzewicz2019trapped}) while the fidelity of the three-spin coupling obtained here is at best $\sim 96\%$ for realistic experimental parameters. Furthermore, the GHZ-state preparation only requires two MS gates compared with a single three-spin gate. Therefore, one will not find a greater benefit in using the three-spin gate in creating GHZ states. There are, however, other situations for which a direct implementation of a unitary of the type $e^{-i\alpha (\sigma_i^+\sigma_j^+\sigma_k^++{\rm h.c.})}$ can potentially reduce the time of the operation, and hence the fidelity of the implementation.

An example is the quantum simulation of the U(1) quantum link model presented in this section. The potential of a fully analog implementation of the correlated three-spin dynamics in the quantum link model is a primary motivation for the three-spin scheme of this work, so that digitalization error of the time-evolution operators could be avoided. Nonetheless, while the implementation of the operator $e^{-i\alpha (\sigma_i^+\sigma_j^+\sigma_k^++{\rm h.c.})}$ on three qubits, with a decomposition to MS gates as
\begin{eqnarray}
e^{-i\alpha(\sigma^+_i\sigma^+_j\sigma^+_k+{\rm h.c.})}&=&
e^{\frac{i\alpha}{4}\, \sigma^y_i\sigma^y_j\sigma^x_k} \,  e^{\frac{i\alpha}{4}\, \sigma^y_i\sigma^x_j\sigma^y_k}
\nonumber\\
&& e^{-\frac{i\alpha}{4}\, \sigma^x_i\sigma^x_j\sigma^x_k} \,  e^{\frac{i\alpha}{4}\, \sigma^x_i\sigma^y_j\sigma^y_k}
\nonumber\\
&=& e^{-\frac{i\pi}{4}\, \sigma^y_i\sigma^z_j} \, e^{\frac{i\alpha}{4}\,\sigma^x_j\sigma^x_k} \, e^{\frac{i\pi}{4}\, \sigma^y_i\sigma^z_j}
\nonumber\\
&& e^{-\frac{i\pi}{4}\, \sigma^y_i\sigma^y_j} \, e^{\frac{i\alpha}{4}\, \sigma^z_j\sigma^y_k} \, e^{\frac{i\pi}{4}\, \sigma^y_i\sigma^y_j}
\nonumber\\
&& e^{-\frac{i\pi}{4}\, \sigma^x_i\sigma^y_j} \, e^{-\frac{i\alpha}{4}\, \sigma^z_j\sigma^x_k} \, e^{\frac{i\pi}{4}\, \sigma^x_i\sigma^y_j}
\nonumber\\
&& e^{-\frac{i\pi}{4}\, \sigma^x_i\sigma^x_j} \, e^{\frac{i\alpha}{4}\, \sigma^z_j\sigma^y_k} \, e^{\frac{i\pi}{4}\, \sigma^x_i\sigma^x_j},
\label{eq:three-to-two-decom}
\end{eqnarray}
is exact and incurs no digitalization errors, at least twelve two-qubit MS gates are needed, in addition to a number of single-qubit gates. The fidelity of this operation, given the best MS gates reported, is $\sim 0.995^{12} = 0.942$. Therefore, here a single-shot implementation of the operator in the left-hand side of Eq.~(\ref{eq:three-to-two-decom}) appears to be comparable or even slightly higher in fidelity compared with the digitized scheme, at least for the first few cycles of the pure three-spin evolution, and for certain laser and trap parameters identified in our numerical study.

Performing side-by-side comparison of the fidelity of operations with only MS gates versus using three-spin gates, however, cannot be rigorously done as this stage for the following reasons. First, while our numerical study gives a reliable estimate of the expected performance, it does not include the anticipated experimental imperfections arising from noise (e.g., laser and trap instabilities) and coupling to environment (e.g., heating, and undesired emission and absorption processes). Such effects are increasingly controlled and diminished in experimental platforms, as is evident from achieved high-fidelity MS operations in chains of ions with various sizes, but are nonetheless needed to be taken into account for a fair comparison with the MS gate-based performances. Second, a high degree of optimization and adjustments is applied to devise complex pulse-shaping protocols in the case of a MS gate to minimize the spin-phonon entanglement (i.e., that can occur already at leading order in the Lamb-Dicke parameter) throughout the gate operation~\cite{Zhu06PRL, Roos_2008, Green2015, Leung2018, Blumel21}, hence increasingly high-fidelities reported in recent state-of-the-art demonstrations. Such an optimization has not been investigated in this work and hence the fidelity reported for the three-spin gate corresponds to the simplest implementation. Devising a pulse-shaping protocol for the three-spin gate is far more complex as it requires a simultaneous minimization of both the first- and second-order contributions in the Lamb-Dicke parameter, but can potentially lead to the same improvement in the three-spin gate fidelity as with the MS gate. Such an avenue will be left to future studies.

\section{Conclusion and outlook
\label{sec:conclusion}}
\noindent
Correlated evolution of the (quasi)spin of three ions in a trapped-ion quantum simulator can be achieved via an extension of the standard M{\o}lmer-S{\o}rensen scheme. Explicitly, a resonant combination of two single-sideband and one double-sideband transitions can be combined to effectively couple three spins via virtual phonons. Resonant single- and two-spin effective transitions are also induced, and can either be used to simulate spin systems with competing two- and three-spin interactions in a magnetic field, or if only pure three-spin dynamics is desired, be eliminated with another set of Raman beams driven with carefully tuned Rabi and beatnote frequencies. The effective Hamiltonian in the single-mode approximation (when the lasers are detuned closely to the single and double sidebands of a single mode so that the contribution to the dynamics  from the nearby modes is small) and in the multi-mode approximation are derived, and the leading corrections to the effective picture are qualitatively identified. Given these corrections, a thorough  check of the scheme is presented by conducting a numerical simulation of the exact dynamics (that from all orders in the Lamb-Dicke parameter and including phonons dynamics) in a rotating-wave approximation. This investigation reveals the experimental parameters for Rabi and beatnote frequencies of the Raman beams, as well as the axial trapping frequency, such that with the use of the second drive, the fidelity of generating a GHZ state with only three-spin dynamics reaches $\gtrsim 95 \%$, and single- and two-spin dynamics remains at least $\sim 20$ times slower that the three-spin dynamics. Furthermore, the phonon occupation in participating modes remains far below one for simulations that start in a phonon-less state. These promising analytic and numerical results motivate future experimental implementation of this scheme, and can potentially simplify quantum simulation of spin systems with multi-spin interactions. The important case of a lattice gauge theory is investigated in this context, and the specifications of an upcoming implementation of the quantum link model of the U(1) gauge theory in $1+1$ dimensions within this scheme are detailed. In particular, it is shown through a crude model of interactions that the near-term demonstrations for systems of $\sim 10$ ions can still reveal interesting constrained dynamics of the lattice gauge theory, such as string breaking, despite the anticipated undesired gauge-violating interactions revealed by our numerical simulation. The comparative performance of analog simulation using engineered three-spin dynamics and that  based on solely MS two-spin gates in digitized dynamics depends on the task and model, and a decisive conclusion will need to await experimental benchmarks.

A few extensions and improvements over the scheme developed here can be enumerated, along with potential applications:
\begin{itemize}
\item[$\diamond$]{While our effective Hamiltonian realizes maximally spin-flipping transitions in a given basis, it is easy to see that interactions proportional to $\sigma_i^z \sigma_j^z \sigma_k^z$ can also be realized in an analogous way, extending two-qubit geometric phase gates~\cite{cirac1995quantum, porras2004effective} to correlated three-qubit operations. In fact, a number of parasitic undesired single- and higher-spin interactions coupled to phonon operators will be absent in the three-qubit phase-operation scheme since couplings among the spin operators at the same ion will be absent in the Magnus expansion, i.e., the commutation among $\sigma^z_i$ operators at the same site is vanishing.\footnote{To achieve interactions proportional to $\sigma_z$ only, a bias term must be eliminated~\cite{schneider2012experimental} through setting Raman beam detunings and polarizations properly, as discussed in Refs.~\cite{ Davoudi:2019bhy, britton2012engineered}.}}
\item[$\diamond$]{The three-spin Hamiltonian of this work is engineered with semi-global beams on every triple of nearest-neighbor ions such that with each operation, interactions are of short range. On the other hand, global Raman beams that address all the ions in the chain can generate three-spin couplings with all-to-all interactions. Various coupling profiles for the interactions, both the single-, two-, and three-spin couplings, can be engineered by tuning the Raman beatnote frequencies, see Eqs.~(\ref{eq:H1sigmamulti})-(\ref{eq:H3sigmamulti}), and if individual addressing of the ions is a possibility, by tuning the Rabi frequencies as well. Such an optimization of parameters to achieve certain coupling profiles is customary in MS-based analog simulation schemes, see e.g., Refs.~\cite{korenblit2012quantum, Davoudi:2019bhy, teoh2020machine}. Furthermore, for digital gate-based applications, pulse-shaping techniques such as those applied to generate optimized MS gates~\cite{Zhu06PRL, Roos_2008, Green2015, Leung2018, Blumel21} can be employed in the three-spin scheme to minimize spin-phonon entanglement, and optimize the operation of the associate three-qubit gate, as mentioned before.}
\item[$\diamond$]{Quantum-simulation and quantum-computing possibilities can be expanded in trapped-ion systems by addressing more than two internal hyperfine levels of the ions, hence effectively introducing a higher-spin degree of freedom for encoding information, i.e., a qudit. The success of this encoding has been already demonstrated in Ref.~\cite{senko2015realization} and its scope is analysed further in Ref.~\cite{low2020practical}. In particular, two-spin entangling operations have been realized in such a setting. It is then straightforward to extend the scheme of this work to three-spin transitions, where one or more ions exhibit a higher (quasi)spin. Among the applications of this capability is in approaching the continuum limit of quantum link model which recovers the U(1) lattice gauge theory by increasing the spin of the quantum link~\cite{zache2021achieving}. The fidelity of the effective three-spin dynamics with higher-spin encodings will need to be both numerically and experimentally quantified.}
\item[$\diamond$]{It is interesting to explore the viability of engineering higher-spin effective interactions, following the strategy of this work. Explicitly, resonant transitions involving multi-spin flips, assisted by a number of virtual phonons, can be induced, but such processes will be higher-order contributions in the Lamb-Dicke parameter, and hence exhibit slower dynamics. Furthermore, for effective four-spin interactions, for example, the lower-order resonant transitions will be proportional to both $1/\delta$ and $1/\delta^2$ in the single-mode approximation, and the simple two-drive scheme of this work with almost opposite detunings cannot eliminate these lower-order spin transitions. Nonetheless, more complex schemes can be potentially devised. Perhaps more interestingly is the possibility of inducing resonant spin-spin-phonon transitions by borrowing ideas from the extended MS scheme of this work, hence opening up the possibility of analog quantum simulation of coupled fermion-boson models, including gauge-fermion couplings in lattice gauge theories, extending the recent hybrid analog-digital proposals~\cite{Lamata:2013sta,  mezzacapo2012digital, Davoudi:2021ney}. A detailed theory and numerical investigation will be required to establish if this scheme can be a viable path toward this goal.}
\end{itemize}

While an experimental demonstration of the scheme of this work is the next immediate goal, all these aforementioned directions will be valuable to investigate in future studies. These align with the overarching goal of enhancing and expanding trapped-ion simulator toolkit for applications beyond what is possible today.

\
\

\section*{Acknowledgments}
\noindent
We are grateful to Christopher Monroe for valuable discussions. We acknowledge Andrew Shaw's involvement at early stages of this work.  BA and TG further acknowledge discussions with Maciej Lewenstein. BA acknowledges funding from the European Union’s Horizon 2020 research and innovation programme under the Marie Skłodowska-Curie grant agreement No. 847517. BA and TG acknowledge funding from Fundacio Privada Cellex, Fundacio Mir-Puig, Generalitat de Catalunya (AGAUR Grant No. 2017 SGR1341, CERCA program, QuantumCAT U16-011424, co-funded by ERDF Operational Program of Catalonia 2014-2020), Agencia Estatal de Investigacion (“Severo Ochoa” Center of Excellence CEX2019-000910-S, PlanNational FIDEUA PID2019-106901GB-I00/10.13039/501100011033, FPI), MINECO-EU QUANTERA MAQS (funded by State Research Agency (AEI) PCI2019-111828-2/10.13039/ 501100011033), EU Horizon 2020FET-OPEN OPTOLogic (Grant No 899794), ERC AdGNOQIA, and the National Science Centre, Poland-Symfonia Grant No. 2016/20/W/ST4/00314. TG further acknowledges a fellowship granted by “la Caixa” Foundation (ID100010434, fellowship code LCF/BQ/PI19/11690013). ZD is supported in part by the U.S. Department of Energy's Office of Science Early Career Award, under award no. DE-SC0020271, for theoretical developments for mapping lattice gauge theories to quantum simulators, and by the DOE Office of Science, Office of Advanced Scientific Computing Research (ASCR) Quantum Computing Application Teams program, under fieldwork proposal number ERKJ347, for algorithmic developments for scientific applications of near-term quantum hardware. MH and GP acknowledge support by the DOE Office of Science, Office of Nuclear Physics, under Award no. DE-SC0021143, for designing hardware-specific simulation protocols for applications in nuclear physics. GP is further supported by the Army Research Office (W911NF21P0003), Army Research Lab (W911QX20P0063), and the Office of Naval Research (N00014-20-1-2695). AS is supported by a Chicago Prize Postdoctoral Fellowship in Theoretical Quantum Science.

\bibliography{bibi.bib}
\end{document}